\documentclass[
   aip,
   jcp,
   preprint,
]{revtex4-1}

\usepackage[margin=1in]{geometry}

\usepackage[utf8]{inputenc}
\usepackage[english]{babel}
\usepackage[T1]{fontenc}

\usepackage[usenames,dvipsnames]{xcolor}
\usepackage{booktabs}
\usepackage{tabularx}
\usepackage{array}
\newcolumntype{L}{>{$}l<{$}} 
\usepackage{subcaption}

\usepackage{lipsum}
\usepackage[footnote,nomargin,english,draft]{fixme}
\makeatletter
\def\@lox@prtc{\section*{\@fxlistfixmename}\begingroup\def\@dotsep{4.5}}
\def\@lox@psttc{\endgroup}
\makeatother

\usepackage[overload]{textcase} \usepackage{tabularray}

\usepackage{braket}
\usepackage{amsmath}
\usepackage{bm}
\usepackage{amssymb}
\usepackage{mathtools}
\usepackage{upgreek}

\makeatletter
\newsavebox{\@brx}
\newcommand{\llangle}[1][]{\savebox{\@brx}{\(\m@th{#1\langle}\)}\mathopen{\copy\@brx\mkern2mu\kern-0.9\wd\@brx\usebox{\@brx}}}
\newcommand{\rrangle}[1][]{\savebox{\@brx}{\(\m@th{#1\rangle}\)}\mathclose{\copy\@brx\mkern2mu\kern-0.9\wd\@brx\usebox{\@brx}}}
\makeatother

\newcommand{\boldupright}[1]{\begingroup \let\alpha\upalpha \let\beta\upbeta \let\gamma\upgamma \let\delta\updelta \let\epsilon\upepsilon \let\varepsilon\upvarepsilon \let\zeta\upzeta \let\eta\upeta \let\theta\uptheta \let\vartheta\upvartheta \let\iota\upiota \let\kappa\upkappa \let\lambda\uplambda \let\mu\upmu \let\nu\upnu \let\xi\upxi \let\pi\uppi \let\varpi\upvarpi \let\rho\uprho \let\varrho\upvarrho \let\sigma\upsigma \let\varsigma\upvarsigma \let\tau\uptau \let\upsilon\upupsilon \let\phi\upphi \let\varphi\upvarphi \let\chi\upchi \let\psi\uppsi \let\omega\upomega \let\Gamma\Upgamma \boldsymbol{\mathbf{#1}}\endgroup }

\usepackage{leftidx}

\usepackage[linesnumbered, ruled, vlined]{algorithm2e}

\usepackage[colorlinks]{hyperref}

\usepackage{cleveref}
\usepackage{ulem}

\newcommand{\ocrev}[1]{{#1}}
\newcommand{\ocrevtwo}[1]{{#1}}

\newcommand{\abj}[1]{{#1}}
\newcommand{\revmgh}[1]{{#1}}
\newcommand{\revmads}[1]{{#1}}
\newcommand{\stkout}[1]{\ifmmode\text{\sout{\ensuremath{#1}}}\else\sout{#1}\fi}

 \DeclareMathAlphabet\mathbfcal{OMS}{cmsy}{b}{n}

\newcommand{\rmstateket}[2][]{\ket{#1{\mathrm{#2}}}}
\newcommand{\rmstatebra}[2][]{\bra{#1{\mathrm{#2}}}}

\newcommand{\vac}{\rmstateket{vac}}
\newcommand{\bvac}{\rmstatebra{vac}}
\newcommand{\cre}[2]{{a}^{#1\,\dagger}_{#2^{#1}}}
\newcommand{\ann}[2]{{a}^{#1}_{#2^{#1}}}

\newcommand{\tdvccphase}[1][]{#1{\epsilon}}

\newcommand{\tdvccLagr}[1][]{#1{\mathcal{L}}}

\newcommand{\cretdm}[2]{\tilde{a}^{#1\,\dagger}_{#2^{#1}}}
\newcommand{\anntdm}[2]{\tilde{b}^{#1}_{#2^{#1}}}
\newcommand{\tdmvccbra}{{\bra{\tilde{\Psi}^\prime}}}
\newcommand{\tdmvccket}{{\ket{\tilde{\Psi}}}}
\newcommand{\tdmvccbrabar}{{\bra{\bar{\Psi}^\prime}}}
\newcommand{\tdmvccketbar}{{\ket{\bar{\Psi}}}}
\newcommand{\tdmvccrefket}{{\ket{\tilde{\Phi}}}}
\newcommand{\tdmvccrefbra}{{\bra{\tilde{\Phi}^\prime}}}
\newcommand{\lamp}[4]{l^{#1#2}_{#3^{#1}#4^{#2}}}
\newcommand{\samp}[4]{s^{#1#2}_{#3^{#1}#4^{#2}}}

\newcommand{\hcheckU}[1]{\LeftIdx{U}\check{h}^{mO^m}_{\alpha^m#1^m}}
\newcommand{\hcheckW}[1]{\LeftIdx{W}\check{h}^{mO^m}_{#1^m\alpha^m}}
\newcommand{\SHCU}[1]{\check{h}^{m}_{\alpha^m#1^m}}   \newcommand{\SHCW}[1]{\check{h}^{m}_{#1^m\alpha^m}}   \newcommand{\htilde}[2]{\tilde{h}^{m_0O^{m_0}}_{#1^{m_0}#2^{m_0}}}
\newcommand{\etilde}[3]{\tilde{E}^{#1}_{#2^{#1}#3^{#1}}}
\newcommand{\tautilde}[2]{\tilde{\tau}^{#1}_{#2^{#1}}}
\newcommand{\tautildedd}[2]{\tilde{\tau}^{#1\ddagger}_{#2^{#1}}}
\newcommand{\Csum}{\sum_{O^mO^{m_0}}C^{mm_0}_{O^mO^{m_0}}}
\newcommand{\CsumOne}{\sum_{O^m}C^{m}_{O^m}}
\newcommand{\W}{\mathbf{W}^m_{\!\!\textsc{A}}}
\newcommand{\Wdot}{\dot{\mathbf{W}}^m_{\!\!\textsc{A}}}
\newcommand{\U}{\mathbf{U}^m_{\!\textsc{A}}}
\newcommand{\Udot}{\dot{\mathbf{U}}^m_{\!\textsc{A}}}
\newcommand{\V}{\mathbf{V}^m_{\!\textsc{A}}}
\newcommand{\Vdot}{\dot{\mathbf{V}}^m_{\!\textsc{A}}}
\newcommand{\Fcheckp}{\check{\mathbf{F}}^{\prime m}}
\newcommand{\Fcheck}{\check{\mathbf{F}}^{m}}

\newcommand{\rhoinv}{\left[\bm{\rho}^m\right]^{-1}}
\newcommand{\Felem}[3]{\check{F}^{#1}_{#2^{#1}#3^{#1}}}
\newcommand{\Fpelem}[3]{\check{F}'^{#1}_{#2^{#1}#3^{#1}}}
\newcommand{\epas}[1]{\tilde{E}^{#1}_{p}}
\newcommand{\efor}[1]{\tilde{E}^{#1}_{f}}
\newcommand{\eup}[1]{\tilde{E}^{#1}_{u}}
\newcommand{\edown}[1]{\tilde{E}^{#1}_{d}}
\newcommand{\hpas}[1]{\tilde{h}^{#1}_{p}}
\newcommand{\hfor}[1]{\tilde{h}^{#1}_{f}}
\newcommand{\hup}[1]{\tilde{h}^{#1}_{u}}
\newcommand{\hdown}[1]{\tilde{h}^{#1}_{d}}
\newcommand{\fcheck}[4][]{#1\check{F}^{#2}_{#3^{#2}#4^{#2}}}
\newcommand{\fcheckp}[4][]{#1\check{F}^{\prime #2}_{#3^{#2}#4^{#2}}}
\newcommand{\circH}{\breve{H}^{m}_{2,(\alpha^mv^m)}}
\newcommand{\circHp}{\breve{H}^{\prime m}_{2,(v^m\alpha^m)}}
\newcommand{\circHone}{\breve{H}^{m}_{1,(\alpha^mv^m)}}
\newcommand{\circHonep}{\breve{H}^{\prime m}_{1,(v^m\alpha^m)}}
\newcommand{\circHOneO}{\breve{H}^{mO^m}_{(\alpha^mv^m)}}
\newcommand{\circHOneOp}{\breve{H}^{\prime mO^m}_{(v^m\alpha^m)}}

\newcommand{\Xintermed}[2]{\leftidx{^{m_0O^{m_0}}}{X}{^{#1}_{#2^{#1}}}}
\newcommand{\Wintermed}[2]{\leftidx{^{m_0O^{m_0}}}{W}{^{#1}_{#2^{#1}}}}
\newcommand{\XLintermed}[3]{\leftidx{^{m_0O^{m_0}}}{(XL)}{^{#1#2}_{#3^{#1}}}}
\newcommand{\WSintermed}[3]{\leftidx{^{m_0O^{m_0}}}{(WS)}{^{#1#2}_{#3^{#1}}}}
\newcommand{\Vintermed}[2]{\leftidx{^{m_0O^{m_0}}}{V}{^{#1#2}}}
\newcommand{\LambdaIntermed}[2]{\Lambda^{#1#2}}
\newcommand{\Xiintermed}[3]{\leftidx{^{#2O^{#2}}}{\Xi}{^{#1#2}_{#3^{#1}}}}

\newcommand{\LeftIdx}[1]{{^{#1\mkern-2mu}}}

 \usepackage{acro}

\DeclareAcronym{mctdh}{
   short = MCTDH ,
   long = multiconfiguration time-dependent Hartree ,
}

\DeclareAcronym{mctdhn}{
   short = MCTDH[$n$] ,
   long = systematically truncated multiconfiguration time-dependent Hartree ,
}

\DeclareAcronym{tdh}{
   short = TDH ,
   long = time-dependent Hartree ,
}

\DeclareAcronym{vscf}{
   short = VSCF ,
   long = vibrational self-consistent field ,
}

\DeclareAcronym{eom}{
   short = EOM ,
   long = equation of motion ,
   short-plural-form = EOMs ,
   long-plural-form = equations of motion ,
}

\DeclareAcronym{tdvp}{
   short = TDVP ,
   long = time-dependent variational principle
}

\DeclareAcronym{tdse}{
   short = TDSE ,
   long = time-dependent Schr{\"o}dinger equation ,
}

\DeclareAcronym{cc}{
   short = CC ,
   long = coupled cluster ,
}
\DeclareAcronym{vcc}{
   short = VCC ,
   long = vibrational coupled cluster ,
}

\DeclareAcronym{tdvcc}{
   short = TDVCC ,
   long = time-dependent vibrational coupled cluster ,
}
\DeclareAcronym{tdvci}{
   short = TDVCI ,
   long = time-dependent vibrational configuration interaction ,
}

\DeclareAcronym{vci}{
   short = VCI ,
   long = vibrational configuration interaction ,
}

\DeclareAcronym{ci}{
   short = CI ,
   long = configuration interaction ,
}

\DeclareAcronym{sq}{
   short = SQ ,
   long = second quantization ,
}

\DeclareAcronym{fq}{
   short = FQ ,
   long = first quantization ,
}

\DeclareAcronym{mc}{
   short = MC ,
   long = mode combination ,
}

\DeclareAcronym{mcr}{
   short = MCR ,
   long = mode-combination range ,
   long-plural = s ,
}

\DeclareAcronym{pes}{
   short = PES ,
   long = potential-energy surface,
   long-plural = s,
   short-plural = s
}

\DeclareAcronym{svd}{
   short = SVD ,
   long = singular-value decomposition ,
}
\DeclareAcronym{adga}{
   short = ADGA ,
   long = adaptive density-guided approach ,
}

\DeclareAcronym{rhs}{
   short = RHS ,
   long = right-hand side ,
}

\DeclareAcronym{lhs}{
   short = LHS ,
   long = left-hand side ,
}

\DeclareAcronym{ivr}{
   short = IVR ,
   long = intramolecular vibrational-energy redistribution ,
}

\DeclareAcronym{fft}{
   short = FFT ,
   long = fast Fourier transform ,
}

\DeclareAcronym{spf}{
   short = SPF ,
   long = single-particle function ,
}

\DeclareAcronym{vmcg}{
   short = vMCG ,
   long = variational multi-configurational Gaussian ,
}

\DeclareAcronym{ccs}{
   short = CCS ,
   long = coupled coherent states ,
}

\DeclareAcronym{lls}{
   short = LLS ,
   long = linear least squares ,
}

\DeclareAcronym{itnamo}{
   short = ItNaMo ,
   long = iterative natural modal ,
}

\DeclareAcronym{hf}{
   short = HF ,
   long = Hartree-Fock ,
}

\DeclareAcronym{mcscf}{
   short = MCSCF ,
   long = multi-configurational self-consistent field ,
}

\DeclareAcronym{sop}{
   short = SOP ,
   long = sum-of-products ,
}
\DeclareAcronym{midascpp}{
   short = MidasCpp ,
   long = Molecular Interactions{,} Dynamics And Simulations Chemistry Program Package ,
   tag = abbrev ,
}

\DeclareAcronym{mpi}{
   short = MPI ,
   long = message passing interface ,
}
\DeclareAcronym{ode}{
   short = ODE ,
   long  = ordinary differential equation ,
   short-plural = s ,
   long-plural = s ,
   short-indefinite = an ,
   long-indefinite = an ,
   tag = abbrev ,
}

\DeclareAcronym{bch}{
   short = BCH ,
   long = Baker-Campbell-Hausdorff ,
}

\DeclareAcronym{sr}{
   short = SR ,
   long = single-reference ,
}
\DeclareAcronym{mr}{
   short = MR ,
   long = multi-reference ,
}

\DeclareAcronym{dof}{
   short = DOF ,
   long = degree of freedom ,
   short-plural-form = DOFs ,
   long-plural-form = degrees of freedom ,
}

\DeclareAcronym{hp}{
   short = HP ,
   long = Hartree product ,
}
\DeclareAcronym{tdbvp}{
   short = TDBVP ,
   long  = time-dependent bivariational principle ,
   short-plural = s ,
   long-plural = s ,
   short-indefinite = a ,
   long-indefinite = a ,
   tag = abbrev ,
}
\DeclareAcronym{eleq}{
   short = ELE ,
   long  = Euler-Lagrange equation ,
   short-plural-form = ELEs ,
   long-plural = s ,
   tag = abbrev ,
}
\DeclareAcronym{mrcc}{
   short = MRCC ,
   long = multi-reference coupled cluster ,
}

\DeclareAcronym{tdfvci}{
   short = TDFVCI ,
   long = time-dependent full vibrational configuration interaction ,
}

\DeclareAcronym{tdevcc}{
   short = TDEVCC ,
   long  = time-dependent extended vibrational coupled cluster ,
   short-plural = s ,
   long-plural = s ,
   short-indefinite = a ,
   long-indefinite = a ,
   tag = abbrev ,
}

\DeclareAcronym{holc}{
   short = HOLC ,
   long = hybrid optimized and localized vibrational coordinate ,
}
\DeclareAcronym{acf}{
   short = ACF ,
   long = autocorrelation function ,
}
\DeclareAcronym{fwhm}{
   short = FWHM ,
   long  = full width at half maximum ,
   short-plural = s ,
   long-plural = full widths at half maxima ,
   short-indefinite = an ,
   long-indefinite = a ,
   tag = abbrev ,
}
\DeclareAcronym{tdmvcc}{
   short = TDMVCC ,
   long = time-dependent vibrational coupled cluster with time-dependent modals ,
}
\DeclareAcronym{tdmvcc2}{
   short = TDMVCC[2]/H2 ,
   long = time-dependent vibrational coupled cluster with time-dependent modals at the two-mode coupled level,
}
\DeclareAcronym{pah}{
   short = PAH ,
   long = polyaromatic hydrocarbon ,
   short-plural = s ,
}
\DeclareAcronym{fab}{
   short = FAB ,
   long = full active basis ,
}
\DeclareAcronym{md}{
   short = MD,
   long  = molecular dynamics,
   short-plural =   ,
   long-plural =   ,
   short-indefinite = a ,
   long-indefinite = a ,
   tag = abbrev ,
}
\DeclareAcronym{qmd}{
   short = QMD,
   long  = quantum molecular dynamics,
   short-plural =   ,
   long-plural =   ,
   short-indefinite = a ,
   long-indefinite = a ,
   tag = abbrev ,
}

\newcommand{\au}{Department of Chemistry, Aarhus University, \\ Langelandsgade 140, DK--8000 Aarhus C, Denmark}
\newcommand{\upo}{Dipartimento di Scienze e Innovazione Tecnologica, Universit\`a del Piemonte Orientale (UPO), Via T. Michel 11, 15100 Alessandria, Italy}

\begin{document}

\title{Efficient time-dependent vibrational coupled cluster computations with time-dependent basis sets
   at the two-mode coupling level: full and hybrid TDMVCC[2]}

\author{Andreas Buchgraitz Jensen}
\email{buchgraitz@chem.au.dk}
\affiliation{\au}

\author{Mads Greisen H\o jlund}
\affiliation{\au}

\author{Alberto Zoccante}
\affiliation{\upo}

\author{Niels Kristian Madsen}
\affiliation{\au}

\author{Ove Christiansen}
\email{ove@chem.au.dk}
\affiliation{\au}

\hypersetup{pdftitle={tdmvcc2h2 implementation}}
\hypersetup{pdfauthor={A.~B.~Jensen, et al.}}
\hypersetup{bookmarks=true,bookmarksopen=true}

\date{\today}

\begin{abstract}
\section*{Abstract}
The computation of \revmgh{the} nuclear quantum dynamics of molecules is challenging, requiring
both accuracy and efficiency to be applicable to systems of interest. 
Recently, theories \revmgh{have} been developed for 
employing time-dependent basis functions \ocrev{(denoted modals) }
with vibrational coupled cluster theory (TDMVCC).
\revmgh{The TDMVCC} method was introduced along with a pilot implementation\revmgh{,} which illustrated good accuracy in benchmark computations.
In this \revmgh{paper} we report an efficient implementation of TDMVCC 
\revmgh{covering the case}
where \revmgh{the wave function} and Hamiltonian \revmgh{contain} up to two-mode couplings. 
After a careful regrouping of \revmgh{terms,} the wave function can be propagated
with a cubic computational scaling with respect to the number of degrees of freedom. 
We discuss the use of a restricted set of active one-mode basis functions \revmgh{for each} mode, 
\revmgh{as well as} two interesting limits: i) the use of a full active basis where the variational 
modal determination amounts essentially to the variational determination of a time-dependent
reference state for the cluster expansion\revmgh{; and} ii) the use of a single function as active basis 
for some degrees of freedom. The latter \revmgh{case defines} a hybrid TDMVCC/TDH approach 
which can obtain an even lower computational scaling.
The resulting computational scaling for hybrid and full 
TDMVCC[2] are illustrated \revmgh{for polyaromatic hydrocarbons} (PAHs) with up to 264 modes.
\revmgh{Finally,} computations on \revmgh{the internal} vibrational redistribution on
benzoic acid \revmgh{(39 modes)} are used to show the faster convergence of 
TDMVCC/TDH hybrid computations towards TDMVCC compared to simple neglect of 
some degrees of freedom.
\end{abstract}

\maketitle

\acresetall

\section{Introduction} \label{sec::introduction}

Molecular dynamics research covers a broad spectrum from classical \ac{md} approaches to fully quantum descriptions. 
Classical \ac{md} methods have the advantage of being easily applicable to large systems using simple coordinates,
but purely classical \ac{md} \revmgh{falls} short in capturing quantum effects that are crucial at the atomic level. 
\ocrev{Quantum-classical trajectory methods incorporate quantum corrections and such 
methods have seen significant progress through the years\revmads{;} see Refs.~\citenum{ple_routine_2023} and \citenum{markland_nuclear_2018} 
and references therein. 
In this \revmads{paper} we will, however,
focus entirely on wave \revmads{function} methods. 
Quantum} dynamical wave function methods are typically computationally expensive, \revmads{i.e. they often exhibit exponential scaling of the cost with respect to the system size.}
\revmads{While this limitation is not universal,}
many \revmads{methods} struggle to handle systems with numerous coupling terms 
and can only treat a limited number of degrees of freedom \revmgh{efficiently}.

Recently, \ac{vcc}\cite{christiansen_vcc_2004} methods \revmgh{have} been extended to the time-dependent domain 
as \ac{tdvcc}\cite{hansen_timedependent_2019,madsen_general_timedependent_2020}, 
\revmgh{thus offering a new approach for solving the \ac{tdse} \revmads{for molecular motion} numerically.}
\ocrev{Time-dependent coupled cluster methods have previously
been discussed in many contexts\cite{arponen1983vpl,kvaal_ab_2012,pigg_time-dependent_2012,hagen_coupled-cluster_2014,sverdrup_ofstad_time-dependent_nodate}, 
and there \revmads{has been} considerable recent activity in 
time-dependent coupled cluster theory for electrons. \cite{kvaal_ab_2012,pedersen_symplectic_2019,sato_communication:_2018,nascimento_linear_2016,pedersen_interpretation_2021,wang_accelerating_2022} We shall in the remainder
of this work focus exclusively on nuclear motion, although a number of formal aspects are similar.} 
\ac{tdvcc} has demonstrated its potential as an efficient method for simulating time-dependent phenomena with attractive features
with respect to accuracy, convergence in configuration space, and separability \revmgh{properties},
but it has also certain drawbacks. In \revmgh{the} initial formulation it relies on both a time-independent basis set and a time-independent reference state, 
which \revmads{restricts the applicability of the method.}
\ocrev{The use of cluster amplitudes to describe effects \revmads{that could be captured in a more compact way by}
time-dependent basis sets} and reference states leads to reduced computational efficiency.
To address these limitations, the concept of \ac{tdmvcc}\cite{madsen_time-dependent_2020,hojlund_bivariational_2022,hojlund_exponential_basis_2023} has been proposed even more recently. 
\ocrev{The term modal refers simply to one-mode basis functions, similar to the single-particle orbital basis of electronic structure theory from which many-particle basis functions are constructed as products.} 
\ac{tdmvcc} aims to combine the advantages of \ac{tdvcc} with the use of \revmgh{a} time-dependent adaptive basis in the spirit of the \ac{mctdh}\cite{meyer_mctdh_1990,beck_multiconfiguration_2000} method. 
\ac{mctdh} has proven highly successful in describing quantum dynamics by utilizing a \revmgh{small, adaptive} active space, providing accurate results at a reasonable computational cost. 
However, \ac{mctdh} still suffers from exponential scaling with the number of modes in the system.
\revmads{While \ac{mctdh} offers a fully correlated description, the much simpler \ac{tdh} method amounts to a mean-field treatment of the dynamics. In fact, \ac{tdh} can
be viewed as a limiting case of \ac{mctdh} where each mode has only a single active modal. This simplifying circumstance reduces the computational scaling to quadratic }\ocrevtwo{for two-body Hamiltonians
in an efficient implementation.\cite{madsen_exponential_2018,ls_vscf09}}
To overcome the scaling issue of \ac{mctdh}, efforts have been made to achieve polynomial scaling by 
\revmads{multilayer schemes (ML-MCTDH)\cite{wan03:1289,wan09:024114,man08:164116,ven11:044135,wang_multilayer_2015} or by truncating the configuration space 
\cite{selected_mctdh, wodraszka_using_2016, wodraszka_systematically_2017, larsson_dynamical_2017,madsen_mrmctdh_2020,madsen_systematic_2020}.
The latter direction includes methods such as MCTDH$[n]$\cite{madsen_systematic_2020} and MR-MCTDH$[n]$\cite{madsen_mrmctdh_2020}
where the wave function is constructed in terms of excitations from a reference configuration. This approach
is similar to how we will truncate the excitation space of \ac{tdmvcc} in this paper.
}\revmads{Other directions include}
\revmads{Gaussian-based} \ac{mctdh} variants (G-MCTDH) 
and other multi-configurational methods. 
\cite{burghardt1999aat,burghardt2008mqd,worth_full_2003,Richings15,shalashilin_multidimensional_2001,James16,curchod18,romer_gaussian-based_2013,eisenbrandt_gaussian-based_2018,greene_tensor-train_2017,dutra_quantum_2020,murakami_accurate_2018,saller_quantum_2017,baiardi_large-scale_2019}
See \revmads{also} Ref.~\citenum{curchod18} for a review \revmads{of multiple spawning methods}.
\ac{tdmvcc} seeks to incorporate the benefits of \ac{mctdh} and the \ac{tdvcc} Ansatz. 
By introducing a small active space of time-dependent basis functions as in \ac{mctdh} and leveraging the fast convergence in configuration 
space offered by the \revmads{exponentially} parameterized \ac{vcc} methodology, 
truncated \ac{tdmvcc} variants \revmads{present} a potential solution to the exponential scaling problem. 
Still, the development of \ac{tdmvcc} is in its infancy and much further research is necessary to fully explore its capabilities. 

The previous pilot implementation of \ac{tdmvcc} could not establish 
the true low-order polynomial computational scaling of \ac{tdmvcc} methods, 
as it relied on very general, but \revmgh{exponentially} scaling steps. It was made 
solely with the aim \revmgh{of investigating} accuracy and convergence in simple benchmarks. 
In this paper, we present an efficient implementation of the
\revmgh{TDMVCC method covering the case where the wave function and Hamiltonian contain up to two-mode couplings (\revmads{henceforth denoted} TDMVCC[2]).
The two-mode coupling case has a manageable number of terms that can be programmed by hand,
something that has been utilized in similar papers in time-independent\cite{seidler2008tfc,zoccante154101} and time-dependent\cite{hansen_timedependent_2019} cases.}
We demonstrate how \revmgh{the TDMVCC[2]} method can be implemented to efficiently 
handle many-mode computations with many-term Hamiltonians, achieving a computational scaling of third order \revmgh{with respect} to the system size. 
\revmgh{In particular,} we address the challenge of efficiently computing the \ac{tdmvcc} mean fields. 
By manually deriving and implementing all the necessary terms, we establish a highly efficient code and lay the groundwork for future general implementations of \ac{tdmvcc}.
In addition, we discuss two interesting limits of using 
a restricted set of active one-mode basis functions \revmads{for each} mode: 
i) the use of a full active basis where the variational
modal determination amounts essentially to the variational determination of a time-dependent
reference state for the cluster expansion\revmgh{; and} ii) the use of a single function as active basis
for some degrees of freedom. The latter \revmgh{case defines a} hybrid TDMVCC/TDH approach\revmgh{, where some modes are treated at the TDH level}.
This approach can obtain an even lower computational scaling, where \revmads{the} \acp{eom} \revmgh{for the \ac{tdh} modes} are similar to 
those of the simple \ac{tdh} theory, only with some \revmgh{additional} mean-field contributions due to the \ac{tdmvcc} modes.
This work thereby \revmads{initiates} a new way of dealing with large systems
with different levels of \ac{tdmvcc} \revmgh{theory}. 

The structure of \revmgh{the} paper is as follows. In \revmgh{Sec.} \ref{sec::theory}, we introduce the necessary theory for 
the \ac{tdmvcc} method and derive the expressions in a form suitable for efficient implementation.
In \revmgh{Sec.} \ref{sec::num_results}, we present numerical results that highlight the features of the two-mode implementation, including computations on  
\acp{pah} with up to 264 modes, \revmgh{as well as} TDMVCC and TDMVCC/TDH 
hybrid computations on benzoic acid.
Finally, in \revmgh{Sec.} \ref{sec::summary}, we provide a summary of our findings and offer an outlook for future research.

 \section{Theory} \label{sec::theory}

\subsection{The many-mode second quantization formulation for biorthogonal \revmads{bases}}
The \ac{tdmvcc} method is formulated using the \ac{sq} framework introduced in Ref.~\citenum{christiansen2004sqf}. 
In an extension to Ref.~\citenum{christiansen2004sqf} and following
Refs.~\citenum{madsen_time-dependent_2020} and \citenum{hojlund_bivariational_2022}, we are working in a biorthogonal framework with the main points repeated in this paper for completeness. 
The biorthogonality is indicated by using $\tilde{a}^\dagger$ and $\tilde{b}$ for creation and annihilation operators, respectively. The tilde is used to denote the time-dependence of the operators.

\revmads{Relative to the} vacuum state, $\vac$, single-mode creation and annihilation operators act as,
\begin{align}
   &\cretdm{m}{s}\vac = \ket{\tilde{\varphi}^m_{s^m}}, \\
   &\bvac \anntdm{m}{r} = \bra{\tilde{\varphi}'^m_{r^m}}.
\end{align}
\revmads{In first quantization coordinate representation, the right-hand sides would be written as $\tilde{\varphi}^m_{s^m}(q_m,t)$ and $\tilde{\varphi}'^m_{r^m}(q_m,t)$.}
The explicit time-dependence will usually be omitted in the following.
One-mode basis functions will generally be denoted modals (in analogy to orbitals for electronic structure theory). 

To describe the biorthogonality of the basis, we define the commutator relations
\begin{equation}
   \Big[\anntdm{m}{r}, \cretdm{m'}{s}\Big] = \delta_{mm'}\delta_{r^m s^{m'}}, \quad \Big[\cretdm{m}{r}, \cretdm{m'}{s}\Big] = \Big[\anntdm{m}{r}, \anntdm{m'}{s}\Big] = 0.
\end{equation}

The time-dependent biorthogonal creation and annihilation operators 
will be defined through a linear combination of time-independent orthonormal creation ($\cre{m}{\alpha}$) and annihilation ($\ann{m}{\alpha}$) operators with a set of time-dependent expansion coefficients:
\begin{align}
   &\cretdm{m}{p} = \sum_{\alpha^m} \cre{m}{\alpha} U^m_{\alpha^mp^m}\label{eq::cretdm}, \\
   &\anntdm{m}{p} = \sum_{\alpha^m} W^m_{p^m\alpha^m} \ann{m}{\alpha}\label{eq::anntdm}.
\end{align}
We use an index convention with Greek letters $\alpha^m$ for the primitive set of time-independent modals
and roman letters $p^m$, $q^m$, $r^m$, $s^m$ for generic time-dependent modals. 
The full set of time-dependent modals is divided into 
an active set of modals indexed by $u^m$, $v^m$, $w^m$ and a set of secondary
modals indexed by $x^m$, $y^m$. The active set of modals are those actually present
in the wave functions. \abj{The number of time-independent modals will later be denoted $N$ while the number of time-dependent modals will be denoted $A$.}

The biorthonormality of the time-dependent modals is captured by the modal expansion coefficients\revmads{,}
\begin{equation}\label{eq::WU_identity}
   \sum_{\alpha^m} W^m_{p^m\alpha^m}U^m_{\alpha^mq^m} = \delta_{p^mq^m} \quad \Leftrightarrow \quad \mathbf{W}^m\mathbf{U}^m = \mathbf{I}^m,
\end{equation}
where we have used matrix notation in the latter equation.

From the creation and annihilation operators we define the shift operator as
\begin{equation}\label{eq::shift}
   \etilde{m}{r}{s} = \cretdm{m}{r}\anntdm{m}{s}.
\end{equation}
This type of operator is used extensively in Sec.~\ref{ssec::2_mode_impl} either as written above in Eq.~\eqref{eq::shift} or as the one-mode excitation and deexcitation operators,
\begin{align}
   &\tautilde{m}{a} = \etilde{m}{a}{i}, \\
   &\tautildedd{m}{a} = \etilde{m}{i}{a}.
\end{align}
Here, excitation and deexcitation refer to a reference \revmads{Hartree product} which is given as a string of $M$ creation operators acting on the vacuum:
\begin{equation}\label{eq::modals}
   \ket{\tilde \Phi} = \prod_{m=1}^M \cretdm{m}{i}\vac\revmads{.}
\end{equation} 
\revmads{Here, $M$} denotes the number of modes.
The index $i^m$ is reserved for the occupied active modal of mode $m$
in the reference state while $a^m$, $b^m$, $c^m$ denote unoccupied \revmads{(virtual)} active modals. 

We will at times use the compound index $\mu$ to mean some collection of mode and modal indices defining a general excitation. 
Other single \revmads{Hartree product} ket or bra states can, for example, be obtained as
\begin{align}
   &\ket{\mu} = \tilde{\tau}_\mu \tdmvccrefket\revmads{,} \\
   &\bra{\mu^\prime} = \tdmvccrefbra \tilde{\tau}^\ddagger_\mu .
\end{align}
In addition, we will use the nomenclature that $\mu_j$ denotes a \textit{j}-mode excitation. In our case the two-mode excitations, $\mu_2$, are of particular interest. 
Such excitations are constructed as products of one-mode excitations, i.e. 
\begin{equation}\label{eq::twomodeexci}
   {\tilde \tau}_{\mu_2} = {\tilde \tau}^{m_1m_2}_{a^{m_1}a^{m_2}} = \etilde{m_1}{a}{i} \etilde{m_2}{a}{i}.
\end{equation}

\subsection{The \ac{tdmvcc} method} \label{ssec::tdmvcc}
To obtain an approximation to the time-dependent wave function which exactly solves the \ac{tdse} in well-defined limits, we employ the \ac{tdmvcc} method as introduced in Ref.~\citenum{madsen_time-dependent_2020}.
This section will give a brief overview of the general scheme of the method as well as 
a summary of the important equations one needs to implement. 

The \ac{tdmvcc} bra and ket \foreignlanguage{ngerman}{Ansätze} are \revmads{given} as
\revmads{\begin{align}
   \tdmvccbrabar &= \tdmvccbra e^{i\tdvccphase}= \tdmvccrefbra (1 + L) e^{-T}e^{i\tdvccphase} , \\
   \tdmvccketbar &= e^{-i\tdvccphase}\tdmvccket = e^{-i\tdvccphase}e^T\tdmvccrefket,
\end{align}
}
\revmads{Here,} the \textit{T} and \textit{L} operators are defined as
\revmads{\begin{align}
   T &= \sum_\mu s_\mu \tilde{\tau}_\mu ,\\
   L &= \sum_\mu l_\mu \tilde{\tau}_\mu^\ddagger\revmads{.}
\end{align}
}
Following the \ac{tdbvp}\revmads{\cite{arponen1983vpl,kvaal_ab_2012}}, one defines a Lagrangian
\begin{equation}\label{eq::tdmvcclagrangian}
   \tdvccLagr = \tdmvccbrabar \left(i\frac{\partial}{\partial t} - \hat{H}\right) \tdmvccketbar ,
\end{equation}
and solves the corresponding \acp{eleq} for all parameters. Taking biorthogonality constraints into account, one arrives at\cite{madsen_time-dependent_2020}
\begin{align}
   \dot{\tdvccphase} &= \tdmvccrefbra \big(\hat{H} - \hat{g}\big) \tdmvccket, \\
   \dot{\mathbf{s}} &= -i\bm{\omega}^{\hat{H} - \hat{g}} , \\
   \dot{\mathbf{l}} &= i\bm{\eta}^{\hat{H} - \hat{g}} , \\
   i\Wdot &= -\tilde{\mathbf{g}}^m \, \W - \rhoinv \Fcheckp \, \mathbf{Q}^m, \\
   i\Udot &= \U \, \tilde{\mathbf{g}}^m - \mathbf{Q}^m \, \Fcheck \rhoinv.
\end{align}
\revmads{Here,} $\mathbf{s}$ and $\mathbf{l}$ are vectors containing the amplitudes, $s_\mu$ and $l_\mu$. Several other definitions are given in Table~\ref{tab::variable_def}.

\begin{table}[ht!]
\centering
\caption{Definition of variables for TDMVCC in \revmads{the general case} and in the specific two-mode case where $T \rightarrow T_2$, $L \rightarrow L_2$, $\mu \rightarrow \mu_2$ and $\hat{H}=\hat{H}_1+\hat{H}_2$. 
For the mean-field matrices we have introduced the semi-locked Hamiltonian operators of Eqs. 
\revmads{(\ref{eq::H_lockgen}, \ref{eq::H_lockgenp}, \ref{eq::circHOneO}--\ref{eq::Hp_lock})}.
The two-mode specific expressions for the $\mathbf{u}^m$ vector, density matrix and the mean-field matrices are also expanded upon in \revmads{Secs.} \ref{sssec::u_vector}, \ref{sssec::density_matrix} and \ref{sssec::mean_field_equations}\revmads{,} respectively. 
}
\label{tab::variable_def}
\begin{tabular}{l|l|l}
\hline\hline
\revmads{Symbol}              & General definition                   & \revmads{Two}-mode specific expression                                         \\ \hline
${\omega}^{\hat{H}}_{\mu}$ & $\braket{\mu^\prime | e^{-T}\hat{H}e^{T} | \tilde{\Phi}}$ & $\braket{\mu_2^\prime | \hat{H}_2 + [\hat{H}_1+\hat{H}_2, T_2] + \frac{1}{2}[[\hat{H}_2, T_2], T_2] \tdmvccrefket}$  \\
${\eta}^{\hat{H}}_{\mu}$   & $\tdmvccbra [\hat{H},{\tilde \tau}_{\mu}] \tdmvccket$ & $\tdmvccrefbra (1+L_2) ([\hat{H}_1 + \hat{H}_2,{\tilde \tau}_{\mu_2}] + [[\hat{H}_2,{\tilde \tau}_{\mu_2}], T_2]) \tdmvccrefket$ \\
$\rho^m_{w^mv^m}$    & $\tdmvccbra \etilde{m}{v}{w} \tdmvccket $           & $\tdmvccrefbra (1+L_2)(\etilde{m}{v}{w} + [\etilde{m}{v}{w}, T_2]) \tdmvccrefket $ \\
$\check{F}^m_{\alpha^mv^m}$           & $\tdmvccbra \cretdm{m}{v}[\ann{m}{\alpha}, \hat{H}]\tdmvccket$  & $\big[\LeftIdx{U}\check{\mathbf{H}}^m_1\bm{\rho}^m\big]_{\alpha^mv^m} + \tdmvccbra \circH \tdmvccket$  \\
$\check{F}^{\prime m}_{v^m\alpha^m}$  & $\tdmvccbra [\hat{H}, \cre{m}{\alpha}]\anntdm{m}{v} \tdmvccket$ & $\big[\bm{\rho}^m\,\,\LeftIdx{W}\check{\mathbf{H}}^m_1\big]_{v^m\alpha^m} + \tdmvccbra \circHp \tdmvccket$  \\
${u}^m_{a^m}$        & $\tilde{F}^{'m}_{a^mi^m} - \tilde{F}^{m}_{a^mi^m} + \text{more terms  }$  & $\tilde{F}^{'m}_{a^mi^m} - \tilde{F}^{m}_{a^mi^m}$ \\
\hline\hline
\end{tabular}
\end{table}

The projection matrix $\mathbf{Q}^m = \revmads{\mathbf{1}^m} - \mathbf{U}^m\mathbf{W}^m$ plays the role of projecting onto the secondary space. 
The $\tilde{\mathbf{g}}^m$ matrix, also termed the constraint matrix, is related to the time-derivative of the creation
and annihilation \revmads{operators} as 
\begin{equation}
   \left[\anntdm{m}{u}, \dot{\tilde{a}}^{m\dagger}_{v^m}\right] = -i\tilde{g}^m_{u^mv^m}, \quad \left[\dot{\tilde{b}}^m_{u^m}, \cretdm{m}{v}\right] = i\tilde{g}^m_{u^mv^m}.
\end{equation}
As discussed in Ref.~\citenum{madsen_time-dependent_2020}, the \ac{tdbvp} leads to the following $\tilde{\mathbf{g}}^m$ matrix structure\revmads{:}
\begin{align}
   \tilde{\mathbf{g}}^m &= \begin{bmatrix}
                              0 & (\leftidx{^d}{\tilde{\mathbf{g}}}{^m})^T \\
                              \leftidx{^u}{\tilde{\mathbf{g}}}{^m} & 0
                           \end{bmatrix}\revmads{.}
\end{align}
The vectors \revmads{$\leftidx{^d}{\tilde{\mathbf{g}}}{^m}$} and $\leftidx{^u}{\tilde{\mathbf{g}}}{^m}$ are found by solving \revmads{two sets of} linear equations:
\begin{align}
   \mathbf{Z}^m\, \leftidx{^d}{\tilde{\mathbf{g}}}{^m} &= \bm{\eta}^{\hat{H},m}, \label{eq::g_down} \\
   -(\mathbf{Z}^m)^T\, \leftidx{^u}{\tilde{\mathbf{g}}}{^m} &= \mathbf{u}^{m}, \label{eq::g_up}
\end{align}
\revmads{Here, we have defined} $Z^m_{a^mb^m} = \delta_{a^mb^m}\rho^m_{i^mi^m} - \rho^m_{a^mb^m}$. The density matrix \revmads{$\bm{\rho}^m$ and the $\mathbf{u}^m$ vector} are discussed in detail for the two-mode case in Secs. \ref{sssec::density_matrix} and \ref{sssec::u_vector}, respectively.

The \ac{tdmvcc} theory reduces to \ac{tdvcc} with a time-independent basis when $\mathbf{W}^m$ and $\mathbf{U}^m$ are kept time-independent and $\mathbf{g}^m$ consequently zero, as described in Refs.~\citenum{hansen_timedependent_2019} and \citenum{madsen_general_timedependent_2020}. In particular, the same structure of the \acp{eom} is found for the phase, $\tdvccphase$, cluster amplitudes, $\mathbf{s}$, and left amplitudes, $\mathbf{l}$. 
A significant difference though is the necessity of the constraint matrix, $\mathbf{g}^m$, and determining the time-evolving modals which introduces the need for computing mean-field matrices.

If there is no secondary space, i.e. if all modals are active, the equations for \revmads{$\Udot$ and $\Wdot$} simplify significantly as the $\mathbf{Q}^m$ matrix becomes the null matrix effectively zeroing the second term in the two \acp{eom}.
The first column of the mean-field matrices are still needed for computing the $\mathbf{u}^m$ vector that is \revmads{used} 
for computing the variationally optimal $\tilde{\mathbf{g}}^m$\revmads{; Eqs.~\eqref{eq::g_down} and \eqref{eq::g_up}}.
In \revmads{Sec.~}\ref{sssec::full_active_basis} we will discuss specifically this limit we will call the \ac{fab}.

\subsection{Polar parameterization of the modals} \label{ssec::polar}

In Ref.~\citenum{hojlund_bivariational_2022} we described how instabilities can occur in the \ac{tdmvcc} \acp{eom} when employing a secondary modal basis (i.e. when the active basis is not full), 
and that such instabilities can also occur for other \revmads{types of wave functions} when a biorthogonal basis is used.
This phenomenon was analyzed in detail for general bivariational time-dependent wave functions, and it was shown that the issue occurs because the bra and ket modal bases tend to span different spaces when given enough time in a numerical computation.
A solution was proposed based on the reparameterization of the modals through polar decomposition.
Fully variational \acp{eom} were derived and subsequently modified in order to lock the bra and ket modal bases together such that they span the same space at any given time. This restricted polar representation corresponds to a modal parameterization of the form
\begin{subequations} \label{eq:polar_param}
\begin{align}
   \U &= \mathbf{V}_{\!\textsc{A}}^m \mathbf{P}^m, \\
   \W &= (\mathbf{P}^m)^{-1} (\mathbf{V}_{\!\textsc{A}}^m)^{\dagger},
\end{align}
\end{subequations}
where $\V$ is semi-unitary (has orthonormal columns) and $\mathbf{P}^m$ is Hermitian.
Eqs.~\eqref{eq:polar_param} clearly imply that the columns of $\U$ and $\W$
span the same space. Although the modified \acp{eom} are not strictly variational,
we found an attractive combination of stability and accuracy.
The \acp{eom} for the polar modal parameters are given by
\begin{subequations}
\begin{align}
   i \Vdot &= \V \tilde{\mathbf{g}}^{\prime m} + \mathbf{Q}^{\prime m} \mathbf{X}^m, \\
   i \dot{\mathbf{P}}^m     &= \mathbf{P}^m     \tilde{\mathbf{g}}^{\prime \prime m}.
\end{align}
\end{subequations}
The necessary matrices are all calculated from known quantities in a straightforward manner, using $\mathbb{H}$ and $\mathbb{A}$ to denote the \revmads{Hermitian} and anti-\revmads{Hermitian parts} of a matrix:
\begin{subequations}
\begin{align}
   \mathbf{Q}^{\prime m} &= \mathbf{1} - \V (\V)^{\dagger}, \\
   \mathbf{X}^m          &= \tfrac{1}{2} \big( \check{\mathbf{F}}^m  (\bm{\rho}^m)^{-1} (\mathbf{P}^m)^{-1} + ( \mathbf{P} (\bm{\rho}^m)^{-1} \check{\mathbf{F}}^{\prime m} )^{\dagger}\big), \\
   \tilde{\mathbf{g}}^{\prime m} &= \mathbb{H}(\bar{\mathbf{g}}^m) + \mathbf{T}^m \big(\mathbf{\Gamma}^{m} \circ ((\mathbf{T}^m)^{\dagger} \mathbb{A}(\bar{\mathbf{g}}^m)  \mathbf{T}^m)\big)   (\mathbf{T}^m)^{\dagger}, \\
   \tilde{\mathbf{g}}^{\prime \prime m} &= \mathbb{A}(\bar{\mathbf{g}}^m) + \mathbf{T}^m \big(\mathbf{\Gamma}^{m} \circ ((\mathbf{T}^m)^{\dagger} \mathbb{A}(\bar{\mathbf{g}}^m)  \mathbf{T}^m)\big)   (\mathbf{T}^m)^{\dagger}.
\end{align}
\end{subequations}
Here, we have diagonalized $\mathbf{P}^m$ as $\mathbf{P}^m = \mathbf{T} \bm{\epsilon} \mathbf{T}^{\dagger}$ and
defined the auxiliary matrices $\bar{\mathbf{g}}^m$ and $\mathbf{\Gamma}^m$ as
\begin{subequations}
\begin{align}
   \bar{\mathbf{g}}^m &= \mathbf{P}^m \tilde{\mathbf{g}}^m  (\mathbf{P}^m)^{-1}, \\
   \gamma_{p^m q^m}^m &= \frac{-\epsilon^m_p + \epsilon^m_q}{\epsilon^m_p + \epsilon^m_q}.
\end{align}
\end{subequations}
The restricted polar representation requires the original $\mathbf{g}^m$ matrices and a few extra steps that scale only linearly \revmads{with respect to} the number of modes. This extra cost can safely be considered negligible.
We will accordingly employ the restricted polar representation throughout 
without further discussion of its computational cost and features.

\subsection{Two-mode TDMVCC implementation} \label{ssec::2_mode_impl}
In the two-mode specific case we truncate the cluster expansion to include at most \revmads{two-mode} excitations. \revmads{As mentioned, one-mode} excitations are redundant
and can be excluded as derived in \revmads{Sec.~}II.C.3 \revmads{of} Ref.~\citenum{madsen_time-dependent_2020}. 
\revmads{This means that unlike in TDVCC$[2]$ (with static modals), we do not need one-mode excitations in TDMVCC$[2]$, since
these degrees of freedom are covered by the time-dependent modals.}
TDMVCC[2] is defined by \revmads{letting} $T \rightarrow T_2$, $L \rightarrow L_2$ \revmads{and} $\mu \rightarrow \mu_2$.
In this study we further restrict ourselves to Hamiltonians containing at most two-mode couplings, i.e. $\hat{H}=\hat{H}_1+\hat{H}_2$, where $\hat{H}_j$ contains only \textit{j}-mode couplings. 
To be specific, let us write the cluster operators in detail:
\begin{align}
T&= \sum_{m_1<m_2} T^{m_1m_2} = \sum_{\mu_2} s_{\mu_2} \tilde{\tau}_{\mu_2} = \sum_{m_1<m_2} \sum_{a^{m_1}a^{m_2}} \samp{m_1}{m_2}{a}{a} \tilde{\tau}^{m_1m_2}_{a^{m_1}a^{m_2}}, \label{eq::t2exp}  \\
L&=\sum_{m_1<m_2} L^{m_1m_2} = \sum_{\mu_2} l_{\mu_2} \tilde{\tau}_{\mu_2}^\ddagger = \sum_{m_1<m_2} \sum_{a^{m_1}a^{m_2}}\lamp{m_1}{m_2}{a}{a} \tilde{\tau}^{m_1m_2\ddagger}_{a^{m_1}a^{m_2}}. \label{eq::l2exp} 
\end{align}

Note that the summations over modes in the $T_2$ and $L_2$ operators runs over unique pairs of modes.
In many summations later in this paper we will use a notation allowing us to utilize permutation
symmetry, i.e. we set $T^{m_1m_2}=T^{m_2m_1}$ and 
${s}^{m_1m_2}_{a^{m_1}a^{m_2}} = {s}^{m_2m_1}_{a^{m_2}a^{m_1}}$ 
and \revmads{similarly} for $L$. 
\revmads{The reasoning is that excitations in two modes $m_1$ and $m_2$ commute, as the mode indices are different by definition.}
\revmads{We can use this in detailed expressions including commutation, e.g.}
\begin{align}\label{eq::t2exps} 
[\etilde{m}{p}{q},T] &= \sum_{m_0>m} [\etilde{m}{p}{q},T^{mm_0}] + \sum_{m_0<m} [\etilde{m}{p}{q},T^{m_0m}]  \nonumber \\ 
                     &= \sum_{m_0>m} [\etilde{m}{p}{q},T^{mm_0}] + \sum_{m_0<m} [\etilde{m}{p}{q},T^{mm_0}]  \nonumber \\ 
                     &= \sum_{m_0 \neq m} [\etilde{m}{p}{q},T^{mm_0}]  \nonumber \\ 
                     &= \sum_{m_0 \neq m} \sum_{a^{m}a^{m_0}} \samp{m}{m_0}{a}{a}[\etilde{m}{p}{q},\etilde{m}{a}{i}] \etilde{m_0}{a}{i}.
\end{align}

We express the Hamiltonian in a \ac{sop} format with factors
containing one-mode operators,  $\hat{h}^{mO^m}$,
where the $O^m$ index refers to a specific operator. 
Correspondingly, we have one-mode integrals of this 
operator in the time-dependent basis denoted as $\tilde{h}^{mO^m}_{p^mq^m}$.
We can write the $ \hat{H}_1$ and $ \hat{H}_2$ operators as 
\begin{align}
   \hat{H}_1 & = \sum_m \hat{H}_1^m = \sum_m \CsumOne \hat{h}^{mO^m} = \sum_{m} \CsumOne \sum_{p^mq^m}\tilde{h}^{mO^m}_{p^mq^m}\etilde{m}{p}{q} , \label{eq::H1} \\
   \hat{H}_2 &= \sum_{m < m_0} \hat{H}_2^{mm_0} = \sum_{m < m_0} \Csum \hat{h}^{mO^m} \hat{h}^{m_0O^{m_0}} \nonumber \\ 
   & = \sum_{m < m_0}\Csum \sum_{\substack{p^{m}q^{m} \\ t^{m_0}u^{m_0}}} \tilde{h}^{mO^m}_{p^{m}q^{m}} \htilde{t}{u} \etilde{m}{p}{q}\etilde{m_0}{t}{u} \revmads{.}  \label{eq::H2}
\end{align}
Here, $C^{m}_{O^m}$ and $C^{mm_0}_{O^mO^{m_0}}$ are possible fitting coefficients. 

The two-mode implementation presented in this paper is obtained by explicitly deriving and hand-coding every single non-zero term. 
The overall idea and implementation follows that of Refs.~\citenum{hansen_timedependent_2019}, \citenum{seidler2008tfc} and \citenum{zoccante154101}, 
but the mean fields contain other types of terms. As the \acp{eom} for the $s_{\mu_2}$ and $l_{\mu_2}$ amplitudes presented in 
\revmads{Sec.}~\ref{ssec::tdmvcc} are equivalent to the ones derived explicitly in Ref.~\citenum{hansen_timedependent_2019} 
(apart from the lack of the $\hat{g}$ operator), we re-use the $\bm{\omega}^{\hat{H}}$ and $\bm{\eta}^{\hat{H}}$ transformers, 
as well as the phase evaluation algorithm. The only terms that are left for us to implement are thus the \revmads{constraint} contributions, the modal derivatives, 
and in particular the efficient evaluation of the \revmads{$\mathbf{u}^m$} vector, the density matrix \revmads{($\bm{\rho}^m$)} and the mean-field matrices \revmads{($\Fcheck$ and $\Fcheckp$)}. 
Of these terms, the mean-field matrices are the most complex to evaluate. They contain many demanding terms, and would be strongly time dominating unless 
care is taken in reducing operation count and scaling by analysis and use of intermediates. 
\revmads{Naively}, the computation of the mean-field matrices would scale as $M^4$, but this can be reduced to $M^3$ as explored in later sections.

Including only two-mode excitations in the excitation space and in the Hamiltonian greatly reduces the number of terms from thousands to tens.
A first indication of this is given later in \revmads{Tables} \ref{tab::mean_field_contribs} and \ref{tab::mean_field_prime_contribs}.

\subsubsection{\revmads{The density} matrix} \label{sssec::density_matrix}
We will start out with an explicit derivation of the TDMVCC[2] density matrix, as it showcases general ideas that can be applied to many of the following equations.
Starting from the definition of the density matrix, we simply expand the bra and ket states and then apply a \revmads{\ac{bch} expansion} which terminates after only two terms, because the $T$ and $L$ operators only contain two-mode excitations/deexcitations.
This procedure yields
\begin{align}
   \rho^{m}_{w^m v^m} &= \tdmvccbra\etilde{m}{v}{w}\tdmvccket \nonumber\\
   &= \tdmvccrefbra (1+L_2) e^{-T_2}\etilde{m}{v}{w}e^{T_2} \tdmvccrefket \nonumber\\
   &= \tdmvccrefbra \big(1+L_2\big)\big( \etilde{m}{v}{w} + \big[\etilde{m}{v}{w}, T_2\big]\big) \tdmvccrefket \nonumber\\
   &= \delta_{i^mw^m}\delta_{i^mv^m} + \tdmvccrefbra L_2\big[\etilde{m}{v}{w}, T_2\big] \tdmvccrefket.
\end{align}
Specifying the modals as either reference or virtual, only four combinations are possible:
\begin{enumerate}
   \item Passive: ($v^m, w^m) = (i^m, i^m$)
   \item Up: ($v^m, w^m) = (a^m, i^m$)
   \item Down: ($v^m, w^m) = (i^m, a^m$)
   \item Forward: ($v^m, w^m) = (a^m, b^m$).
\end{enumerate}
It is rather easy to see that the up and down elements vanish as the shift operator becomes either an excitation or deexcitation operator, i.e. $\rho^m_{a^mi^m} = \rho^m_{i^ma^m} = 0$. The remaining cases; passive and forward give non-zero contributions. The passive indices results in
\begin{align}\label{eq::rho_i_i}
   \rho^m_{i^mi^m} &= 1 + \tdmvccrefbra L_2\left[\tilde{E}^{m}_{i^mi^m}, T_2\right] \tdmvccrefket \nonumber\\
   &= 1 + \sum_{m_0 \neq m}\sum_{a^m b^{m_0}} \samp{m}{m_0}{a}{b} \tdmvccrefbra L_2 \left[\etilde{m}{i}{i}, \tilde{\tau}^{mm_0}_{a^mb^{m_0}} \right] \tdmvccrefket \nonumber\\
   &= 1 - \sum_{m_0 \neq m}\sum_{a^m b^{m_0}} \samp{m}{m_0}{a}{b} \tdmvccrefbra L_2 \tilde{\tau}^{mm_0}_{a^mb^{m_0}} \etilde{m}{i}{i}\tdmvccrefket \nonumber\\
   &= 1 - \sum_{m_0 \neq m}\sum_{a^m b^{m_0}} \samp{m}{m_0}{a}{b} \lamp{m}{m_0}{a}{b}.
\end{align}
Following the same logic we can get an expression for the forward contribution, where we obtain a sum over matrix products given as
\begin{equation}\label{eq::density_virtual_contrib}
   \rho^m_{b^ma^m} = \sum_{m_0 \neq m} \sum_{c^{m_0}} \samp{m}{m_0}{b}{c}\lamp{m}{m_0}{a}{c}.
\end{equation}

\subsubsection{\revmads{The} $\mathbf{u}^m$ vector} \label{sssec::u_vector}
Another important quantity to compute is the $\mathbf{u}^m$ vector; see Eq.~\revmads{(A29)} in Ref.~\citenum{madsen_time-dependent_2020}. 
Utilizing the fact that $\mu$ is strictly a two-mode excitation ($\mu_2$) simplifies this expression significantly. Equation \revmads{(A29)} reads
\begin{equation}\label{eq::u_vector}
   u^m_{a^m} = \tilde{F}^{'m}_{a^mi^m} - \tilde{F}^{m}_{a^mi^m} + \sum_\mu \left(\tdmvccbra \big[\tautildedd{m}{a}, \tilde{\tau}_\mu\big] \tdmvccket (i\dot{s}_\mu) + (i\dot{l}_\mu)\bra{\tilde{\mu}^{\prime}}e^{-T}\tautildedd{m}{a}\tdmvccket\right).
\end{equation}
\revmads{Assuming that the fully transformed mean-field matrices as well as $\dot{\mathbf{l}}$ and $\dot{\mathbf{s}}$}
are calculated and saved elsewhere, we are left with the summation over $\mu$ and write this out with explicit two-mode cluster operators as
\begin{equation}
   u^m_{a^m} \leftarrow \sum_{\mu_2} \left(\tdmvccrefbra (1+L_2)e^{-T_2}\big[\tautildedd{m}{a}, \tilde{\tau}_{\mu_2}\big] e^{T_2}\tdmvccrefket (i\dot{s}_{\mu_2}) + (i\dot{l}_{\mu_2})\bra{\tilde{\mu}_2^{\prime}}e^{-T_2}\tautildedd{m}{a}e^{T_2}\tdmvccrefket\right).
\end{equation}
The two terms are expanded using the \revmads{\ac{bch} expansion} and examined one at a time. The \revmads{\ac{bch} expansion} in the first term evaluates to
\begin{equation}
   e^{-T_2}\big[\tautildedd{m}{a}, \tilde{\tau}_{\mu_2}\big] e^{T_2} = \big[\tautildedd{m}{a}, \tilde{\tau}_{\mu_2}\big] + \Big[\big[\tautildedd{m}{a}, \tilde{\tau}_{\mu_2}\big], T_2\Big] + \bigg[\Big[\big[\tautildedd{m}{a}, \tilde{\tau}_{\mu_2}\big], T_2\Big], T_2\bigg].
\end{equation}
It is seen that the resulting three terms contains an even number of excitations (2, 4, 6) due to $\tilde{\tau}_{\mu_2}$ and $T_2$, and a single deexcitation due to $\tautildedd{m}{a}$. In the full expression we also have the $(1+L_2)$ term containing 0 and 2 deexcitations. This means that in total we have an even number of excitations (2, 4, 6) and an odd number of deexcitations (1, 3) resulting in orthogonal bra and ket states. The first term in the $\mu_2$ sum in Eq.~\eqref{eq::u_vector} therefore always contributes 0 to the $\mathbf{u}^m$ vector.
The same procedure can be applied to the second term, again resulting in a vanishing contribution
\revmads{\begin{equation}
   \bra{\tilde{\mu}_2^{\prime}}e^{-T_2}\tautildedd{m}{a}e^{T_2}\tdmvccrefket = \bra{\tilde{\mu}_2^{\prime}}\left(\tautildedd{m}{a} + \big[\tautildedd{m}{a},T_2\big] + \Big[\big[\tautildedd{m}{a}, T_2\big], T_2\Big] \right)\tdmvccrefket = 0 . 
\end{equation}
}Thus, the $u^m_{a^m}$ vector elements for the two-mode couplings case is simply the difference of the first columns in the mean-field matrices, i.e.
\begin{equation}\label{eq::u_vec_final}
   u^m_{a^m} = \tilde{F}^{'m}_{a^mi^m} - \tilde{F}^{m}_{a^mi^m}.
\end{equation}
Again, this encourages an efficient implementation of the mean-field matrices which will be presented in \revmads{Sec.~}\ref{sssec::mean_field_equations}.

\subsubsection{Mean-field equations} \label{sssec::mean_field_equations}
As mentioned in \revmads{Sec.~}\ref{ssec::tdmvcc} an efficient implementation of the mean-field matrices is key to an efficient implementation of the \ac{tdmvcc} method. 
We will thus elaborate on the specifics of the equations for the mean fields in this and the next section. 
The \revmads{half-transformed} mean fields defined in Ref.~\citenum{madsen_time-dependent_2020} are given as
\begin{align}
   \check{F}^{m}_{\alpha^m v^m} &= \tdmvccbra \cretdm{m}{v}{}[\ann{m}{\alpha}, \hat{H} ] \tdmvccket,\label{eq::i_mode_f}  \\
   \check{F}^{\prime m}_{v^m\alpha^m} &= \tdmvccbra [\hat{H}, \cre{m}{\alpha} ]\anntdm{m}{v} \tdmvccket \label{eq::i_mode_fp}.
\end{align}

To obtain computationally convenient expressions for the two-mode part we rewrite the primitive creation and annihilation operators in the biorthogonal basis using Eqs.~\eqref{eq::cretdm}, \eqref{eq::anntdm} and \eqref{eq::WU_identity}, giving the following two expressions
\begin{align}
   &\ann{m}{\alpha} = \sum_{p^m}U^m_{\alpha^mp^m}\tilde{b}^m_{p^m}, \label{eq::ann_prim}\\
   &\cre{m}{\alpha} = \sum_{p^m}\tilde{a}^{m\dagger}_{p^m}W^m_{p^m\alpha^m} \label{eq::cre_prim}.
\end{align}
This means that 
\begin{align}
   &[\ann{m}{\alpha},\etilde{m}{p}{q}] = U^m_{\alpha^mp^m} \tilde{b}^m_{q^m}, \label{eq::commut_a_e} \\
   &[\etilde{m}{p}{q},\cre{m}{\alpha}] = W^m_{q^m\alpha^m} \cretdm{m}{p}. \label{eq::commut_e_ad}
\end{align}
We can therefore rewrite the essential commutator part of the mean-field expressions using
\begin{align}
   \circHOneO = \cretdm{m}{v} [\ann{m}{\alpha},\hat{h}^{mO^m}] 
   & = \cretdm{m}{v} \sum_{p^mq^m} {\tilde h}^{mO^m}_{p^mq^m} [\ann{m}{\alpha},\etilde{m}{p}{q}] \nonumber \\
   & = \cretdm{m}{v} \sum_{p^mq^m} {\tilde h}^{mO^m}_{p^mq^m} U^m_{\alpha^mp^m} \tilde{b}^m_{q^m} \nonumber \\ 
   & = \sum_{q^m} \hcheckU{q} \etilde{m}{v}{q} \label{eq::H_lockgen}
\end{align}
\begin{align}
\circHOneOp = [\hat{h}^{mO^m},\cretdm{m}{\alpha}]\ann{m}{v} 
   & = \sum_{p^m} \hcheckW{p} \etilde{m}{p}{v} \label{eq::H_lockgenp}
\end{align}
Here, we have introduced one-mode integrals that are transformed halfway into the primitive basis:
\begin{align}
   \hcheckU{q}  &= \sum_{p^m}U^m_{\alpha^mp^m}\tilde{h}^{mO^m}_{p^mq^m}, \label{eq::hchecku1}\\
   \hcheckW{w}  &= \sum_{q^m}W^m_{q^m\alpha^m}\tilde{h}^{mO^m}_{p^mq^m}  \label{eq::hcheckw1}.
\end{align}

\revmads{Now, if Eqs.~\eqref{eq::H1}, \eqref{eq::H2}, \eqref{eq::commut_a_e} and \eqref{eq::commut_e_ad} are inserted into Eqs.~\eqref{eq::i_mode_f} and \eqref{eq::i_mode_fp},}
we can write the one- and two-mode Hamiltonian contributions to the mean fields as expectation values of effective operators. We thus obtain 
\begin{align}
   &\fcheck[\LeftIdx{}]{m}{\alpha}{v} = \tdmvccbra \circHone  \tdmvccket + \tdmvccbra \circH  \tdmvccket,   \\ 
   &\fcheckp[\LeftIdx{}]{m}{v}{\alpha} = \tdmvccbra \circHonep \tdmvccket + \tdmvccbra \circHp \tdmvccket,
\end{align} 
with the mode-locked operators defined as
\begin{align}
   \circHone  & = \CsumOne \circHOneO,  \label{eq::circHOneO} \\ 
   \circHonep  &= \CsumOne \circHOneOp, \label{eq::circHOneOp}\\
   \circH  & = \sum_{\substack{m_0 < m \\ m_0 > m}} \Csum \circHOneO  \tilde{h}^{m_0O^{m_0}},\label{eq::H_lock} \\
   \circHp  &= \sum_{\substack{m_0 < m \\ m_0 > m}} \Csum \circHOneOp \tilde{h}^{m_0O^{m_0}}.\label{eq::Hp_lock}
\end{align}

In accord with Ref.~\citenum{madsen_time-dependent_2020}, the contributions to the mean fields from $\hat{H}_1$ can be computed efficiently from the density matrix as
\begin{align}
   \fcheck[\LeftIdx{1M}]{m}{\alpha}{v} &= \tdmvccbra \circHone  \tdmvccket =  \CsumOne  \sum_{w^m} \hcheckU{w} \rho^{m}_{w^mv^m} = \Big[\LeftIdx{U}\check{\mathbf{H}}^m_1\,\bm{\rho}^m\Big]_{\alpha^mv^m}, \\
   \fcheckp[\LeftIdx{1M}]{m}{v}{\alpha} &= \tdmvccbra \circHonep  \tdmvccket =  \CsumOne  \sum_{w^m} \hcheckW{w} \rho^{m}_{v^mw^m} = \Big[\bm{\rho}^m\,\,\LeftIdx{W}\check{\mathbf{H}}^m_1\Big]_{v^m\alpha^m}.
\end{align}

The two-mode contributions to the mean fields can be written as 
\begin{align}
   \fcheck[\LeftIdx{2M}]{m}{\alpha}{v} &= \sum_{\substack{m_0 < m \\ m_0 > m}} \Csum \sum_{w^m}\hcheckU{w} \sum_{t^{m_0}u^{m_0}}\htilde{t}{u} \tdmvccbra\etilde{m}{v}{w}\etilde{m_0}{t}{u} \tdmvccket,\label{eq::F_start} \\
   \fcheckp[\LeftIdx{2M}]{m}{v}{\alpha} &= \sum_{\substack{m_0 < m \\ m_0 > m}} \Csum \sum_{w^m}\hcheckW{w} \sum_{t^{m_0}u^{m_0}}\htilde{t}{u}  \tdmvccbra\etilde{m}{w}{v}\etilde{m_0}{t}{u}  \tdmvccket.\label{eq::Fm_start}
\end{align}

The two-mode part of the Hamiltonian is much more challenging compared to the one-mode part as we now have two shift operators in our bra-ket expression, giving a set of two-mode density matrix elements. 
Because of this, non-zero \revmads{second-order} terms are now possible in the \revmads{\ac{bch} expansion}. 
We do not want to compute and store the potentially very large set of two-mode density matrix elements but rather seek to 
derive all terms and their explicit form. In \revmads{Tables} \ref{tab::mean_field_contribs} and \ref{tab::mean_field_prime_contribs} we have 
listed all the non-zero contributions to the two mean-field matrices, which can be realized by analyzing 
Eq.~\eqref{eq::F_start} and \eqref{eq::Fm_start} in the same way as discussed in \revmads{Sec.~}\ref{sssec::density_matrix}. 
Table \ref{tab::intermediates} contains a list of intermediates which are used for the computations.

\begin{table}[ht!]
\centering
\caption{Contributions to the mean-field matrix $\Fcheck$. Intermediates listed are defined in Table \ref{tab::intermediates}.}
\label{tab::mean_field_contribs}
\begin{tabular}{llclc}
\hline\hline
No. & $\Fcheck$               & BCH order & Contribution                                  & Intermediates \\ \hline
1   & $\Felem{m}{\alpha}{a}$  & 0 & $\SHCU{i}\hup{m_0} L_2\eup{m}\eup{m_0}$               & W \\
2   & $\Felem{m}{\alpha}{a}$  & 1 & $\SHCU{a}\hpas{m_0}   L_2\efor{m}\epas{m_0}T_2$       & $\Lambda$\\
3   & $\Felem{m}{\alpha}{a}$  & 1 & $\SHCU{i}\hdown{m_0}  L_2\eup{m}\edown{m_0}T_2$       & XL \\
4   & $\Felem{m}{\alpha}{a}$  & 1 & $\SHCU{a}\hfor{m_0}   L_2\efor{m}\efor{m_0}T_2$       & \\ \hline
5   & $\Felem{m}{\alpha}{i}$  & 0 & $\SHCU{i}\hpas{m_0}   \epas{m}\epas{m_0}$             & \\
6   & $\Felem{m}{\alpha}{i}$  & 1 & $\SHCU{a}\hdown{m_0}  \edown{m}\edown{m_0}T_2$        & X \\
7   & $\Felem{m}{\alpha}{i}$  & 1 & $-\SHCU{i}\hpas{m_0}  L_2T_2\epas{m}\epas{m_0}$       & U \\
8   & $\Felem{m}{\alpha}{i}$  & 1 & $\SHCU{i}\hpas{m_0}   L_2\epas{m}\epas{m_0}T_2$       & U \\
9   & $\Felem{m}{\alpha}{i}$  & 1 & $\SHCU{a}\hup{m_0}    L_2\edown{m}\eup{m_0}T_2$       & WS \\
10  & $\Felem{m}{\alpha}{i}$  & 1 & $\SHCU{i}\hfor{m_0}   L_2\epas{m}\efor{m_0}T_2$       & V \\
11  & $\Felem{m}{\alpha}{i}$  & 2 & $\SHCU{a}\hdown{m_0}  L_2\edown{m}\edown{m_0}T_2T_2$  & $\Xi$, U, X \\
12  & $\Felem{m}{\alpha}{i}$  & 2 & $-\SHCU{a}2\hdown{m_0} L_2T_2\edown{m}\edown{m_0}T_2$ & U, X\\
\hline\hline
\end{tabular}
\end{table}

\begin{table}[ht!]
\centering
\caption{Contributions to the mean-field matrix $\Fcheckp$. Names for intermediates listed under column 'I' is defined in Table \ref{tab::intermediates}.}
\label{tab::mean_field_prime_contribs}
\begin{tabular}{llclc}
\hline\hline
No. & $\Fcheckp$              & BCH order & Contribution                                   & I \\ \hline
13  & $\Fpelem{m}{i}{\alpha}$ & 0 & $\SHCW{i}\hpas{m_0}   \epas{m}\epas{m_0}$              & \\
14  & $\Fpelem{m}{i}{\alpha}$ & 0 & $\SHCW{a}\hup{m_0}    L_2\eup{m}\eup{m_0}$             & W \\
15  & $\Fpelem{m}{i}{\alpha}$ & 1 & $-\SHCW{i}\hpas{m_0}  L_2T_2\epas{m}\epas{m_0}$        & U \\
16  & $\Fpelem{m}{i}{\alpha}$ & 1 & $\SHCW{i}\hpas{m_0}   L_2\epas{m}\epas{m_0}T_2$        & U \\
17  & $\Fpelem{m}{i}{\alpha}$ & 1 & $\SHCW{a}\hdown{m_0}  L_2\eup{m}\edown{m_0}T_2$        & XL \\
18  & $\Fpelem{m}{i}{\alpha}$ & 1 & $\SHCW{i}\hfor{m_0}   L_2\epas{m}\efor{m_0}T_2$        & V \\ \hline
19  & $\Fpelem{m}{a}{\alpha}$ & 1 & $\SHCW{i}\hdown{m_0}  \edown{m}\edown{m_0}T_2$         & X \\
20  & $\Fpelem{m}{a}{\alpha}$ & 1 & $\SHCW{a}\hpas{m_0}   L_2\efor{m}\epas{m_0}T_2$        & $\Lambda$\\
21  & $\Fpelem{m}{a}{\alpha}$ & 1 & $\SHCW{i}\hup{m_0}    L_2\edown{m}\eup{m_0}T_2$        & WS \\
22  & $\Fpelem{m}{a}{\alpha}$ & 1 & $\SHCW{a}\hfor{m_0}   L_2\efor{m}\efor{m_0}T_2$        & \\
23  & $\Fpelem{m}{a}{\alpha}$ & 2 & $\SHCW{i}\hdown{m_0}  L_2\edown{m}\edown{m_0}T_2T_2$   & $\Xi$, U, X \\
24  & $\Fpelem{m}{a}{\alpha}$ & 2 & $-\SHCW{i}2\hdown{m_0} L_2T_2\edown{m}\edown{m_0}T_2$  & U, X\\ 
\hline\hline
\end{tabular}
\end{table}

\subsubsection{Intermediates for highly efficient mean-field equations} \label{sssec::intermeds}
If one were to naively implement the mean-field equations as written in \revmads{Sec.}~\ref{sssec::mean_field_equations} the computational cost would \revmads{scale} as $M^4$. 
This would be an order higher than the \ac{tdvcc}[2] and \revmads{time-dependent vibrational configuration interaction with singles and doubles (TDVCI$[2]$)} methods \revmads{(with static modals)}. 
Luckily, we can circumvent this issue by the introduction of intermediates. 
Only a couple of intermediates are strictly necessary for bringing down the scaling exponent, but more have been introduced to bring down the absolute cost as well. 
We \revmads{find} some intermediates introduced in Refs. \citenum{seidler2008tfc} and \citenum{zoccante154101}, but we will repeat them here for completeness and to account for slight changes in definition and/or notation. 
\abj{All intermediates used have been listed in \revmads{Table}~\ref{tab::intermediates} with their definition, storage cost and computational cost. In this table $M^x$ refers to the scaling regarding the number of modes, $A^x$ denotes the scaling \revmads{with respect to the size of the active space}, and $O^x$ is the scaling \revmads{with respect to} the number of operator terms.}
Intermediates $(ls)^{mm_1}_{a^ma^{m_1}}$ and $U^{mm_0}$ are of particular importance, as they allow scaling to be reduced from $M^4$ to $M^3$. 

\begin{table}[ht!]
\centering
\caption{Intermediates used to reduce computational scaling.}
\label{tab::intermediates}
\begin{tabular}{llcc}
   \hline\hline
   Name & Definition & Storage & Computation \\ \hline
   $\Xintermed{m}{a}$ & $\sum_{c^{m_0}}\htilde{i}{c}\samp{m}{m_0}{a}{c}$ & $M^2\abj{A}O$ & $M^2\abj{A}^2O$ \\
   $\Wintermed{m}{a}$ & $\sum_{c^{m_0}}\htilde{c}{i}\lamp{m}{m_0}{a}{c}$ & $M^2\abj{A}O$ & $M^2\abj{A}^2O$ \\
   $\XLintermed{m}{m_0}{a}$ & $\sum_{m_1 \neq \{m, m_0\}}\sum_{c^{m_1}}\Xintermed{m_1}{c}\lamp{m}{m_1}{a}{c}$ & $M^2\abj{A}O$ & $M^3\abj{A}^2O$ \\
   $\WSintermed{m}{m_0}{a}$ & $\sum_{m_1 \neq \{m, m_0\}}\sum_{c^{m_1}}\Wintermed{m_1}{c}\samp{m}{m_1}{a}{c}$ & $M^2\abj{A}O$ & $M^3\abj{A}^2O$ \\
   $(LS)^{m_1m_2}$ & $\sum_{a^{m_1}b^{m_2}}\lamp{m_1}{m_2}{a}{b}\samp{m_1}{m_2}{a}{b}$ & $M^2$ & $M^2\abj{A}^2$ \\
   $U^{mm_0}$ & $\sum_{m_1<m_2 \neq\{m,m_0\}}(LS)^{m_1m_2} - \sum_{m_1 < m_2}(LS)^{m_1m_2}$ & $M^2$ & $M^3$ \\
   $\Vintermed{m}{m_0}$ & $\sum_{m_1 \neq \{m, m_0\}} \sum_{b^{m_1}c^{m_0}d^{m_0}}\lamp{m_1}{m_0}{b}{c}\htilde{c}{d}\samp{m_1}{m_0}{b}{d}$ & $M^2O$ & $M^3\abj{A}^3O$ \\
   $\LambdaIntermed{m}{m_0}$ & $\sum_{m_1\neq \{m, m_0\}}\sum_{c^{m_1}}\lamp{m}{m_1}{a}{c}\samp{m}{m_1}{b}{c}$ & $M^2\abj{A}^2$ & $M^3\abj{A}^3$ \\
   $(ls)^{mm_1}_{a^ma^{m_1}}$ & $\sum_{m_2\neq \{m, m_1\}}\sum_{b^{m_2}}\samp{m}{m_2}{a}{b}\lamp{m_1}{m_2}{a}{b}$ & $M^2\abj{A}^2$ & $M^3\abj{A}^3$ \\
   $\Xiintermed{m}{m_0}{a}$ & $ \sum_{m_1\neq \{m, m_0\}}\left(\sum_{a^{m_1}}\Xintermed{m_1}{a}(ls)^{mm_1}_{a^ma^{m_1}}\right)$ &  &  \\
          & $\phantom{\sum} - \sum_{b^{m_0}}\samp{m}{m_0}{a}{b}\left(\sum_{m_1\neq \{m, m_0\}}\sum_{a^{m_1}}\Xintermed{m_1}{a}\lamp{m_1}{m_0}{a}{b}\right)$ & $M^2\abj{A}O$ & $M^3\abj{A}^2O$ \\
   \hline\hline
\end{tabular}
\end{table}

We will now look at two intermediates in more detail.
The first intermediate, which we will denote \revmads{by} $\Xi$, arises from contributions 11 and 23.
\abj{Expanding contribution 11 and 23 from the compact notation in \revmads{Tables~}\ref{tab::mean_field_contribs} and \ref{tab::mean_field_prime_contribs}, one obtains several terms among which the following leads to the $\Xi$ intermediate:}
\begin{equation}\label{eq::xi_intermed_explicit}
   \sum_{m_1\neq \{m, m_0\}}\sum_{m_2\neq \{m, m_0, m_1\}} \sum_{a^{m_1}b^{m_2}}\Xintermed{m_1}{a}\lamp{m_1}{m_2}{a}{b}\samp{m}{m_2}{a}{b}.
\end{equation}
It can be seen there are 4 mode indices ($m$, $m_0$, $m_1$, $m_2$) giving the $M^4$ scaling. We introduce the following quantity as an intermediate:
\begin{equation}
   (ls)^{mm_1}_{a^ma^{m_1}} = \sum_{m_2\neq m, m_1}\sum_{b^{m_2}}\lamp{m_1}{m_2}{a}{b}\samp{m}{m_2}{a}{b}.
\end{equation}
\revmads{Equation}~\eqref{eq::xi_intermed_explicit} can now be written as
\begin{equation}\label{eq::efficient_3rdbch}
   \sum_{m_1\neq \{m, m_0\}}\left(\sum_{a^{m_1}}\Xintermed{m_1}{a}(ls)^{mm_1}_{a^ma^{m_1}}\right) - \sum_{b^{m_0}}\samp{m}{m_0}{a}{b}\left(\sum_{m_1\neq m, m_0}\sum_{a^{m_1}}\Xintermed{m_1}{a}\lamp{m_1}{m_0}{a}{b}\right).
\end{equation}
Only three mode indices are now present ($m$, $m_0$, $m_1$) giving the scaling of $M^3$. Of course, it is necessary to compute the intermediates, but this also scales as $M^3$ requiring two computations with $M^3$ scaling, which is much more favorable than one computation at $M^4$ scaling.

The second term contributing to the $M^4$ scaling stems from contributions 7, 8, 11, 12, 15, 16, 23 and 24 and appears as
\begin{equation}
   U^{mm_0} = \sum_{m_1 < m_2 \neq \{m, m_0\}} (LS)^{m_1m_2} - \sum_{m_1 < m_2} (LS)^{m_1m_2}.
\end{equation}
Since all the terms in the first summation is contained in the second, the first term is not computed, and it is a matter of checking when the combinations of mode indices result in something not contained in the first summation and then only compute and subtract these.

After introducing these two intermediates \revmads{(along with the remaining intermediates in Table~\ref{tab::intermediates}),} we can write the final mean-field equations to be computed as follows, with only $M^3$ scaling cost,

\begin{align}\label{eq::f_alpha_i}
   \check{F}^m_{\alpha^mi^m} =  & \sum_{\substack{m_0 < m \\ m_0 > m}} \Csum \times  \nonumber \\
                                & \Big\{\hcheckU{i}(\htilde{i}{i} + \Vintermed{m}{m_0} +\htilde{i}{i}U^{mm_0})  \nonumber  \\
                                & \phantom{\{}+\sum_{a^m} \hcheckU{a} \Xintermed{m}{a}(1 + U^{mm_0}) \nonumber   \\
                                & \phantom{\{}+\sum_{a^m}\hcheckU{a}\WSintermed{m}{m_0}{a} \nonumber  \\
                                & \phantom{\{}+\sum_{a^m} \hcheckU{a} \Xiintermed{m}{m_0}{a}\Big\}
\end{align}
\begin{align}\label{eq::f_alpha_a}
   \check{F}^m_{\alpha^ma^m} =  & \sum_{\substack{m_0 < m \\ m_0 > m}} \Csum \times \nonumber   \\
                                & \Big\{\hcheckU{i}\Wintermed{m}{a}  \nonumber  \\
                                & \phantom{\{}+ \htilde{i}{i} \sum_{b^m} \hcheckU{b} \LambdaIntermed{m}{m_0}\nonumber    \\
                                & \phantom{\{}+\hcheckU{i} \XLintermed{m}{m_0}{a}  \nonumber  \\
                                & \phantom{\{}+ \sum_{b^m} \hcheckU{b} \sum_{c^{m_0}d^{m_0}} \lamp{m_0}{m}{c}{a}\tilde{h}^{m_0O^{m_0}}_{c^{m_0}d^{m_0}}\samp{m_0}{m}{d}{b} \big\}
\end{align}
\begin{align}\label{eq::f_i_alpha}
   \check{F}'^m_{i^m\alpha^m} = & \sum_{\substack{m_0 < m \\ m_0 > m}} \Csum \times \nonumber   \\
                                & \Big\{\hcheckW{i} (\htilde{i}{i} + \Vintermed{m}{m_0} + \htilde{i}{i}U^{mm_0}) \nonumber   \\
                                & \phantom{\{}+\sum_{a^m}\hcheckW{a}\Wintermed{m}{a}  \nonumber  \\
                                & \phantom{\{}+ \sum_{a^m}\hcheckW{a}\XLintermed{m}{m_0}{a}\Big\}
\end{align}
\begin{align}\label{eq::f_a_alpha}
   \check{F}'^m_{a^m\alpha^m} = & \sum_{\substack{m_0 < m \\ m_0 > m}} \Csum \times  \nonumber  \\
                                & \Big\{\hcheckW{i}\Xintermed{m}{a}(1 + U^{mm_0})  \nonumber  \\
                                & \phantom{\{}+\htilde{i}{i}\sum_{b^m}\hcheckW{b} \LambdaIntermed{m}{m_0}\nonumber    \\
                                & \phantom{\{}+\hcheckW{i} \WSintermed{m}{m_0}{a}  \nonumber  \\
                                & \phantom{\{}+\sum_{b^m}\hcheckW{b}\sum_{c^{m_0}d^{m_0}}\samp{m}{m_0}{a}{d}\htilde{c}{d}\lamp{m}{m_0}{b}{c} \nonumber   \\
                                & \phantom{\{}+\hcheckW{i} \Xiintermed{m}{m_0}{a} \Big\}
\end{align}

\subsubsection{The full active basis approach}\label{sssec::full_active_basis}

The use of an active space smaller that the full space has been decisive for MCTDH and obviously anticipated
to be similarly advantageous for \ac{tdmvcc}. 
However, it is of course an additional approximation that may give a basis set truncation error,
and it gives rise to additional numerical steps, projections etc.
In the \ac{fab} limit the time-evolving basis has solely the purpose of making the reference optimal, which is obviously a theoretically interesting limit.
For these reasons we consider the efficient implementation of using the full space
as active space to study advantages and disadvantages. 
In the limit where the primitive and time-dependent modal bases are equal in size, the \acp{eom} simplify significantly, and we only need to compute the parts of the mean-field matrices necessary for computing the $\mathbf{u}^m$ vector. We see in Eq.~\eqref{eq::u_vec_final} that this is simply the first column of both the matrices. Expanding the right-hand side of Eq.~\revmads{(112)} from Ref.~\citenum{madsen_time-dependent_2020},
\begin{equation}
   \tilde{F}^{\prime m}_{v^mw^m} - \tilde{F}^{m}_{v^mw^m} = \tdmvccbra [\hat{H}, \etilde{m}{w}{v}] \tdmvccket,
\end{equation}
as was done in the previous sections results in terms with structure similar to the ones in Eq.~\eqref{eq::f_alpha_i} and \eqref{eq::f_i_alpha}, but with fully transformed mean-field matrices rather than \revmads{half-transformed} ones. These two approaches are of course equivalent in the limit of no secondary modal space, but the implementation differs a bit as new efficient intermediates/transformations might become available.

Initially we had hoped for a very fast specialized implementation of this limit, but despite the apparent simplicity of the $\mathbf{u}^m$ vector equations, many expensive terms remain to be computed, and the approach still scales as $M^3$.

\subsection{Hybrid TDMVCC/TDH scheme} \label{ssec::hybrid_scheme}
In this section we present a computational scheme introduced for obtaining an important computational limit which opens up for efficient calculations of large systems. The idea is to apply \ac{tdmvcc} in the usual way for a selected set of important modes and then only have a single active time-dependent modal for the rest. 
Thus, we \revmads{restrict} these modes to stay in the reference modal. In this way we obtain a TDMVCC/TDH hybrid scheme where we do not ignore less important modes, as may be a more standard approach. Instead, they are included in a simplified way which makes the computations significantly faster compared to TDMVCC[2].
We denote the modes with the simplified treatment as TDH modes but emphasize that their mean fields will
be different from the pure TDH mean fields due to the \ac{tdmvcc} modes. 
Thus, one can state that the ket state of the system can be written like
\begin{equation}
   \tdmvccketbar = e^{-i\tdvccphase}e^{T^\mathrm{TDMVCC} }\prod_{m\in \mathrm{MC}^{\mathrm{TDMVCC}}} \cretdm{m}{i}\prod_{m\in \mathrm{MC}^{\mathrm{TDH}}} \cretdm{m}{i} \vac 
\end{equation}
The speedup achieved by TDMVCC/TDH is seen by the fact that we can skip certain parts of the calculations involved in solving the \acp{eom} as these TDH modes do not have any virtual modals. For example, we see this effect in the density matrices and the mean-field matrices. The density matrix for a TDH mode reduces to a 1-by-1 matrix with the element $1$, as can be seen in Eqs. \eqref{eq::rho_i_i} and \eqref{eq::density_virtual_contrib}. 
The mean-field matrices for the TDH modes reduce to a vector with very few contributions as seen in Eqs. \eqref{eq::f_alpha_i} and \eqref{eq::f_i_alpha}, where the intermediates with a virtual modal index does not contribute.

This approach will be explored further with numerical results in \revmads{Sec.~}\ref{ssec::benzoic_acid_results} where we utilize it to do computations on a 39-mode \ac{pes} (coupled at the two-mode level) describing benzoic acid.
We will also denote this method as TDMVCC[2|X] where the X denotes the number of modes propagated at the \ac{tdmvcc} level. The X notation will be used both with absolute numbers, but also with relative quantities indicating a certain percentage of the systems modes which have been propagated at the \ac{tdmvcc} level.
\abj{We note that TDMVCC[2|0] will occasionally be named the TDH limit.}
 \section{Numerical results} \label{sec::num_results}

The described methods have been implemented in the \ac{midascpp}\cite{MidasCpp}. All computational results presented in the following have been obtained with this implementation.

\revmads{We will henceforth use the following nomenclature for the three TDMVCC$[2]$ schemes that are considered in this paper:
i) hybrid TDMVCC$[2]$ (or TDMVCC[2|X]) denotes the scheme where only a subset of modes is treated at the TDMVCC level;
ii) \ac{fab} TDMVCC$[2]$ denotes the implementation that has been optimized to run with an equal number of time-dependent and time-independent modals ($A=N$); and
iii) TDMVCC$[2]$ (with no further qualifiers) denotes the basic scheme where all modes are treated at the TDMVCC level and where the optimized \ac{fab} implementation is not used (even if $A=N$).}

\subsection{Mode scaling} \label{ssec::mode_scaling}
\begin{figure}[ht]
   \centering
   \includegraphics[width=0.95\textwidth]{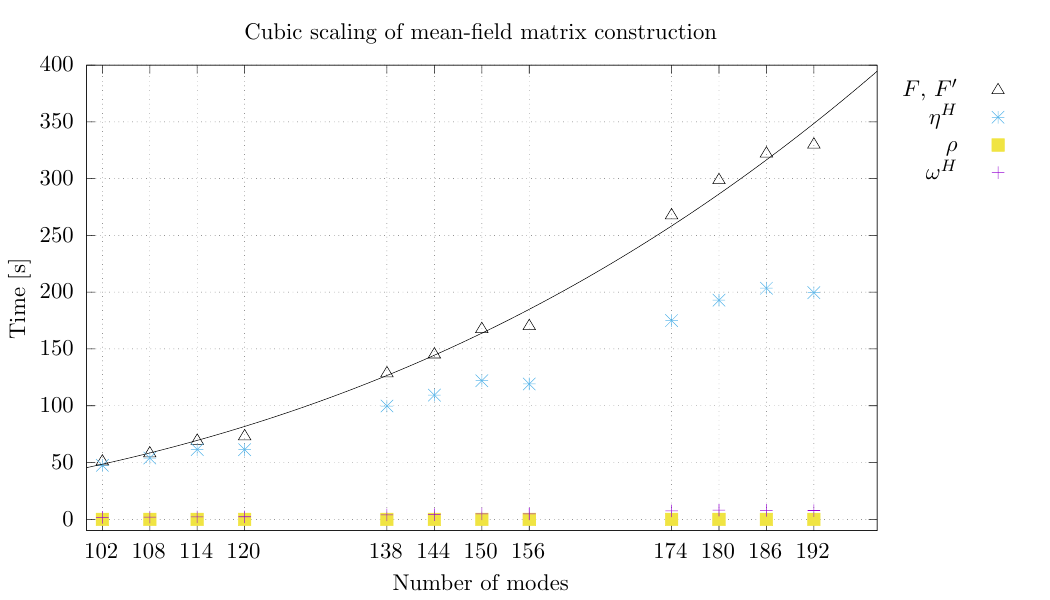}
   \caption{\revmads{Mode-scaling} of the TDMVCC[2] code. \revmads{The time for computing mean fields follows the theoretical $M^3$ scaling (black line) quite well.}
   \revmads{The density matrices are very cheap, as expected.}
   \revmads{The timings for the vectors} $\bm{\omega}^{\hat{H}}$ and $\bm{\eta}^{\hat{H}}$ are also shown for absolute timing perspective.}
   \label{fig::mode_scaling_std_tdmvcc}
\end{figure}

Theoretically we found the scaling of the mean fields with respect to the number of modes in the system to be $M^3$. 
The scaling of the implementation of the mean fields computation has been tested with calculations on 12 \acp{pah} systems with the number of modes ranging 
from 102 to 192\ocrevtwo{ using pre-computed two-mode \acp{pes}\cite{ls_vscf09,seidler2008tfc}}. \revmads{To get representative numbers, we have averaged over 96 evaluations of the mean-field matrices.}
A b-spline basis was used as the primitive basis with bounds at the 10th classical turning point. 
For the modals we used the 5 \revmads{lowest-lying} VSCF modals as our stationary basis and then 5 time-dependent modals for 
Fig.~\ref{fig::mode_scaling_std_tdmvcc}, which is sufficient for investigating the scaling of the method with respect the number of modes.
The \acp{pes} have a \revmads{relatively} large number of terms (in the range 59,000 to 122,000) and are thus representative for real \acp{pes}. 
Obviously, absolute timings would be even more favorable with simple few-term model potentials.

In Fig.~\ref{fig::mode_scaling_std_tdmvcc} the computational times for different part of the \ac{tdmvcc} method has been plotted.
The timings for the $\bm{\omega}^{\hat{H}}$ and $\bm{\eta}^{\hat{H}}$ \revmads{vectors (Refs.~\citenum{seidler2008tfc} and \citenum{zoccante154101})} have also been plotted to give a sense of absolute times.

From Fig. \ref{fig::mode_scaling_std_tdmvcc} we see that a theoretical scaling proportional to $M^{3}$ fits rather nicely with the data for the computation of the total mean-field matrices. We see that the mean-field computations are the major computational task even after our significant optimizations.

\begin{figure}
   \centering
   \includegraphics[width=0.95\textwidth]{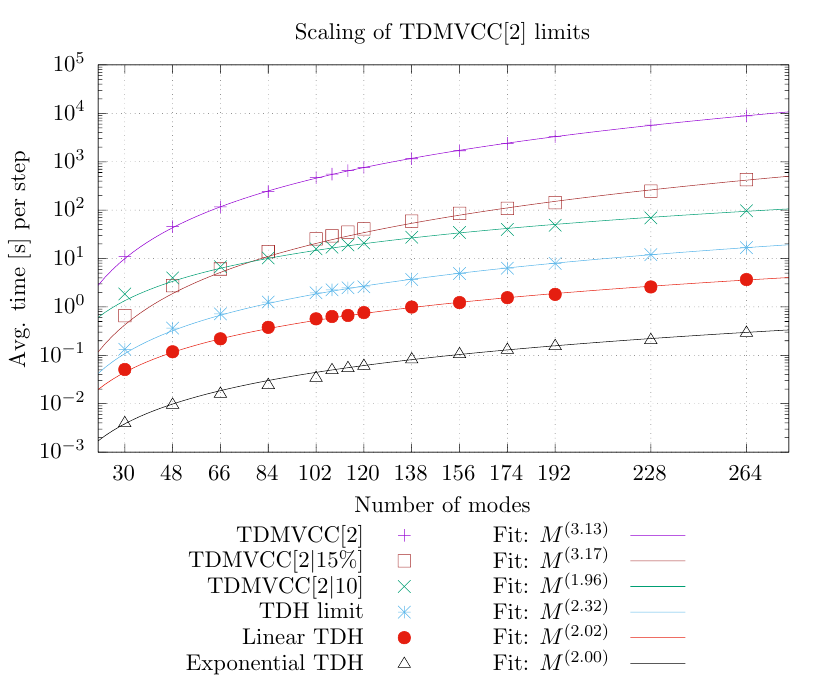}
   \caption{Scaling of \revmads{TDMVCC$[2]$ and several limiting methods. The two data points between 102 and 120 correspond to 108 and 114 modes.} 
   \revmads{TDMVCC[2|10] and TDMVCC[2|15\%] signify that 10 modes or 15\% of the modes, respectively, are treated at the \ac{tdmvcc} level. The remaining modes are propagated at the \ac{tdh} level.}
   }
   \label{fig::ModeScalingMultipleMethods}
\end{figure}

In Fig. \ref{fig::ModeScalingMultipleMethods} we show the scaling of four different computational schemes for \ac{tdmvcc} and for comparison two different versions of TDH. 
For \ac{tdmvcc} we show the approach where all modes are propagated at the \ac{tdmvcc} level, two different hybrid approaches where 
i) \revmads{a fixed number of} 10 modes are propagated at the \ac{tdmvcc} level \revmads{while the rest are kept at the TDH level;} 
ii) \revmads{15\% percent of the modes propagated at the \ac{tdmvcc} level;} and iii) the limit where all modes are propagated at the TDH level. 
These results are also run on PAH systems with a \revmads{range of sizes}. 
Note that the \acp{pes} for these timings have been trimmed, keeping only the \revmads{five lowest-order} coupling terms, 
so that the number of coupling terms are the same for each \ac{mc}. 
This was done as the number of coupling terms in the \acp{pes} \revmads{do} not increase as a simple polynomial as 
a function of the number of modes, giving spurious fitting for the results shown in Fig.~\ref{fig::ModeScalingMultipleMethods}.
From Fig. \ref{fig::ModeScalingMultipleMethods} we see that the TDMVCC[2|X] scheme is computationally very attractive, as the computational times of the \ac{tdmvcc} method can be significantly reduced if the important dynamics is located to a small subset of modes. 
This is especially true if the number of dynamically important modes does not change with respect to the system size. It can be seen that the TDMVCC[2|10] method scales as $M^2$ just as the TDH method. 
That is, the 10 modes propagated at the \ac{tdmvcc} level simply \revmads{result in} a prefactor. If the number of dynamically important modes is not independent on the system size, but still located to a subset of the total modes, we end up with the $M^3$ scaling, as the number of modes propagated at the \ac{tdmvcc} level increases. 
Importantly, this is at a significantly reduced cost compared to propagating all modes at the \ac{tdmvcc} level.

In the limit where all modes are propagated at the TDH level, we see, despite a lot of logical TDMVCC overhead, that the implementation recovers the $M^2$ scaling of TDH\cite{madsen_exponential_2018}. The explicit TDH code clearly runs faster still, with the \revmads{exponentially} parameterized TDH as the clear winner. 
It is an opportunity for future research to implement TDMVCC[2] with exponentially parameterized modals\cite{hojlund_exponential_basis_2023} and recover this efficiency gain. However, very large systems and small TDMVCC[2] domains must be considered for this additional gain to be significant.

\subsection{Basis size scaling} \label{ssec::full_active_basis_mode_scaling}
\begin{figure}[ht]
   \centering
   \includegraphics[width=0.95\textwidth]{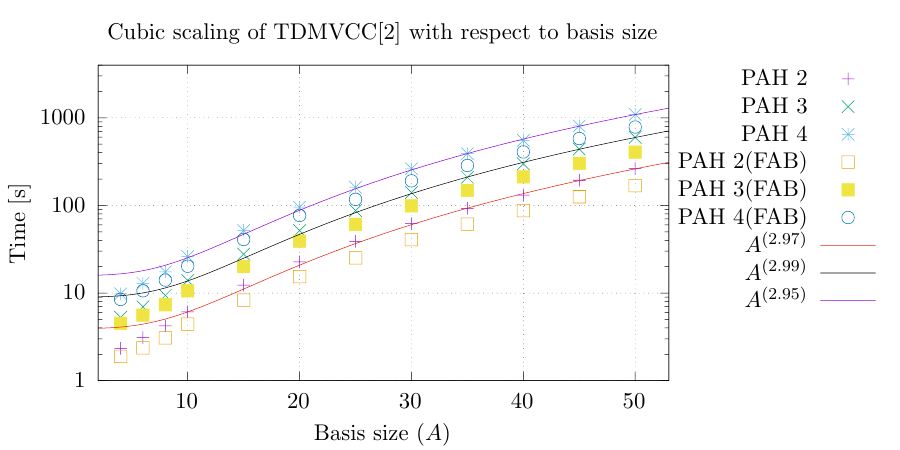}
   \caption{Scaling of \revmads{TDMVCC$[2]$} with respect to the \revmads{size of the} time-dependent active basis \revmads{($A$)}. 
   \revmads{For the TDMVCC$[2]$ implementation, $A$ is varied with $N=50$ fixed. For
   the \ac{fab} implementation, $A=N\leq 50$ is varied.}
   }
   \label{fig::variable_active_basis}
\end{figure}

Here we investigate the scaling of the method with respect to the basis set size. The setup is very similar to the above computations, we have here limited the computations to \ac{pah} systems with 48, 66 and 84 modes, and we only look at the timings for constructing the mean fields as these are the most important factors as shown in Fig. \ref{fig::mode_scaling_std_tdmvcc}.
The basis scaling results have been plotted in Fig. \ref{fig::variable_active_basis}. For \abj{TDMVCC[2]} we have fixed $N=50$ and then varied $A$. \revmads{Additionally, we have plotted} the same results obtained from a \ac{fab} computation ($N=A\leq 50$).

For basis sets larger than 10 we see a reasonable fit to roughly $A^3$. This meets \revmads{our} expectations since there are several terms scaling as $M^3A^3$ as listed in table \ref{tab::intermediates}.

In the \ac{fab} ($A=N$) limit we see a decrease in computing time. 
\revmads{We found the computing time for the TDMVCC[2] implementation to be 1.2--1.6 times slower than the specialized \ac{fab} implementation,}
depending on the system size and the basis size (large basis sets gave larger difference while large systems gave a smaller difference). 
One of course has to hold this up against the quality of the result\revmads{: The} \ac{fab} \revmads{implementation can of course only be used in the $N=A$ limit, which often yields poor results for realistic values of $N$.}
Furthermore, an important point of \ac{tdmvcc} is that it is not necessary to use $N=A$ to obtain good results and often $N$ can be significantly larger than $A$. Therefore, the \ac{fab} option is mainly of interest from a theoretical point of view. We foresee that the restricted \revmads{polar parameterization in combination with a small} active space will be used in applications.

\subsection{IVR for Benzoic acid}\label{ssec::benzoic_acid_results}
To explore further the numerical behavior of TDMVCC[2] and the quality of the TDMVCC[2|X] method we present some results on the benzoic acid molecule \revmads{(39 modes)}. 
The \ac{pes} was obtained using the \ac{adga}\cite{sparta_adaptive_2009} procedure in \ac{midascpp}\cite{MidasCpp} and the GFN2-xTB method\cite{gfn2} 
in the xTB program for \revmads{computing cheap} single points in these exploratory computations. 
\ac{adga} settings are given in the supporting information \revmads{and the final \ac{pes} has 35,956 terms.}
The \revmads{dynamics of the benzoic acid molecule} is initialized by using a \revmads{VSCF state} with the mode describing the acid O--H stretch in the first excited state (described by the $v=1$ modal) computed on an uncoupled \ac{pes} and then letting this state relax in the coupled \ac{pes}.

\begin{figure}
   \centering
   \includegraphics[width=\textwidth]{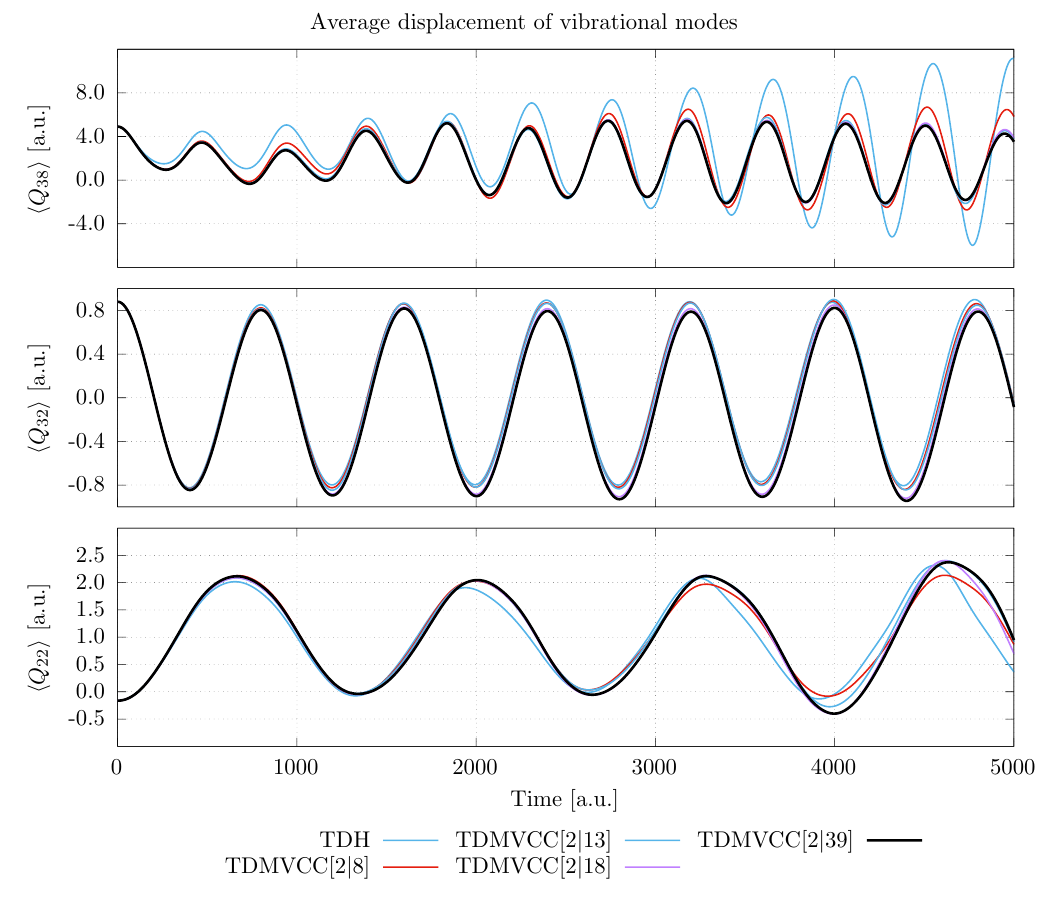}
   \caption{\revmads{Position expectation values for three coordinates of the benzoic acid molecule.}
      $Q_{38}$ describes the O--H stretch. $Q_{32}$ describes the C=O stretch. $Q_{22}$ describes the in-plane bending mode of the O--H group.
      \abj{The TDMVCC[2|39] computation corresponds simply to TDMVCC[2], i.e. all modes are included as TDMVCC modes.}
      }
   \label{fig::tdmvcc2_benzoic_acid_multiplot_occgen_1}
\end{figure}

In Fig. \ref{fig::tdmvcc2_benzoic_acid_multiplot_occgen_1} the expectation value of the displacement coordinate describing the O--H stretch in the acid group ($Q_{38}$) is plotted as a function of time. \revmads{Coordinates $Q_{32}$ and $Q_{22}$ are also shown for comparison.} 
As a reference method we use TDMVCC[2|39] (also \abj{simply named TDMVCC[2]}) where all modes \revmads{are described at the coupled cluster level. Each mode has} an active basis of four time-dependent modals. 
\revmads{We compare the reference calculation with linear \ac{tdh} and with the hybrid TDMVCC[2|X] method with $\mathrm{X} = 8, 13, 18$.}
The modes to be propagated at the \ac{tdmvcc}[2] level were chosen by \revmads{visual inspection (modes dominated by motion localized in the acid group were prioritized).
A table is given in the supporting information listing the specific modes used in the different hybrid schemes.
}Expectation values are \revmads{generally} defined as $\braket{\Lambda | \hat{O} | CC}$ for an operator $\hat{O}$. 
However, for \revmads{the simple case of a one-mode operator $\hat{O}^m$ this expression specializes to} $\sum_{r^ms^m} \tilde{O}^m_{r^ms^m}\rho^m_{s^mr^m}$,
\revmads{which adds no significant additional cost to the computations.}

\revmads{It is clear that increasing X ensures that the expectation values converge towards the 
\ac{tdmvcc}[2] limit, where all modes at treated at the coupled cluster level.}
\revmads{We also note that} if one wants to do computations on a very large system the computational cost can be significantly reduced by the hybrid approach, and the results are qualitatively very similar. Another observation is that the hybrid approach is much better than \ac{tdh}.

\begin{figure}
   \centering
   \includegraphics[width=\textwidth]{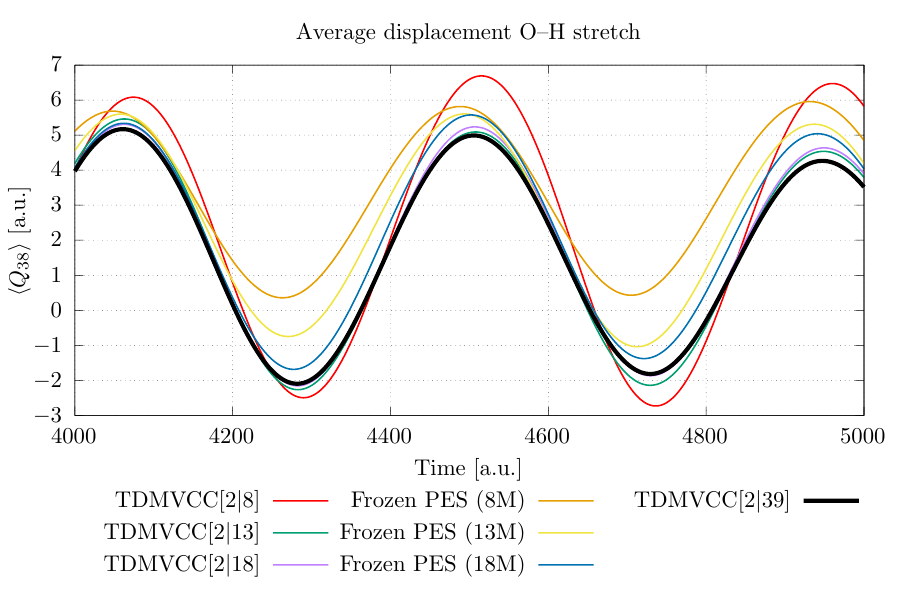}
   \caption{Comparison of the hybrid TDMVCC[2|X] method and reduced dimension \acp{pes} \revmads{for the benzoic acid molecule}.
      \abj{The TDMVCC[2|39] computation corresponds simply to TDMVCC[2], i.e. all modes are included as TDMVCC modes.}
      }
   \label{fig::benzoic_acid_dynamic_zoom_frozen_pes}
\end{figure}

To show that the hybrid method is favorable compared to a computation with frozen modes (i.e. \revmads{a computation in reduced dimensionality}), 
\revmads{we have performed a number of calculations where the less important modes have been frozen completely instead of being included as TDH modes. 
The computational settings are identical to those in Fig. \ref{fig::tdmvcc2_benzoic_acid_multiplot_occgen_1} and the results are shown in Fig. \ref{fig::benzoic_acid_dynamic_zoom_frozen_pes}. 
}\abj{We now only show the results for the O--H stretch (Q38) and only the last 1000 time units to easier distinguish between the results. It is seen that both schemes converges towards the TDMVCC[2] results, but that the hybrid schemes does so with a lower number of TDMVCC modes.}

\abj{Results for the hybrid TDMVCC[2|X] approach are clearly better, but this of course comes with a more time-consuming calculation. In Table \ref{tab::absolute_timings} the approximate timings for several computations shown in Figs. \ref{fig::tdmvcc2_benzoic_acid_multiplot_occgen_1} and \ref{fig::benzoic_acid_dynamic_zoom_frozen_pes} are listed. The propagation \revmads{time is} 5000 a.u. corresponding to about 120 femtoseconds. So far the implementation only allows for serial calculations, which is an obvious area of improvement for future work.}

\begin{table}[ht!]
\centering
\caption{Absolute timings of computations presented in Figs. \ref{fig::tdmvcc2_benzoic_acid_multiplot_occgen_1} and \ref{fig::benzoic_acid_dynamic_zoom_frozen_pes}.}
\label{tab::absolute_timings}
\begin{tabular}{ll}
   \hline\hline
   Calculation name &  Time \\ \hline
   TDH:             &  2 hours \\
   Frozen PES (8M):     &  1.7 hours \\
   Frozen PES (13M):    &  6 hours \\
   Frozen PES (18M):    &  15 hours  \\
   TDMVCC[2|8]:     &  17 hours \\
   TDMVCC[2|13]:    &  1.4 days \\
   TDMVCC[2|18]:    &  1.8 days \\
   TDMVCC[2|39]:    &  7.5 days \\
   \hline\hline
\end{tabular}
\end{table}
 \section{Summary and outlook} \label{sec::summary}

This work presents the first efficient implementation of \revmgh{the \ac{tdmvcc} method}. Focusing
on the two-mode coupling level, all detailed equations were derived and implemented
\revmads{by hand}. 
We showed that the computational cost of TDMVCC[2] can be reduced to cubic scaling
with respect to the number of degrees of freedom by introducing appropriate intermediates. 
Specializations were made, allowing efficient and flexible usage of both
full active basis sets and basis sets limited to only a few active basis function per mode. 
The latter allows for a hybrid TDMVCC/TDH approach 
that has an even lower quadratic computational scaling
as long as the set of \revmads{TDMVCC modes} is kept constant and
only the number of TDH modes is extended. 
The resulting low order scaling was illustrated
for \acp{pah} with up to 264 modes,
and the gain of TDMVCC/TDH hybrid computations compared to simply 
freezing some degrees of freedom 
was illustrated in computations on a 39-mode benzoic acid \ac{pes}.

The ability to perform these high-dimensional computations shows how
TDMVCC may open for accurate quantum dynamical studies
on systems that \revmads{were previously} out of reach. 
The \revmads{hybrid methods introduced} further opens a new line of research where TDMVCC computations
can be tailored to balance accuracy, efficiency, and feasibility in quantum dynamical computations. 

The \revmads{work presented here} is only the first step in making TDMVCC more generally
applicable. In many cases the present two-mode version will not be sufficiently accurate, 
and higher-level TDMVCC and higher-level couplings in the Hamiltonian \revmads{will be} important.
This includes flexible multi-reference schemes along the lines 
of recent \revmads{work on MR-MCTDH$[n]$}\cite{madsen_mrmctdh_2020}.
Implementing such higher-level TDMVCC methods must generally be considered 
\revmads{as a very hard problem} as higher-order TDMVCC \revmgh{methods} with higher-order Hamiltonians \revmads{include numerous terms (on the order of many thousands)}
\cite{madsen_general_timedependent_2020}.
\revmads{For \ac{tdvcc} (with static modals), it has been possible to make a general implementation\cite{madsen_general_timedependent_2020}
with well-defined low-order polynomial scaling of the computational effort by utilizing existing machinery for
automatic derivation, analysis and computation of all terms that appear in the equations. However, \ac{tdmvcc}
contains new types of terms (particularly in the mean fields) that fall outside the scope of our current code base.}
The present work has shown that it is possible and important to construct certain 
appropriate intermediates to obtain low-order computational scaling. 
This work therefore paves the \revmads{way} for and encourages the laborious work 
on general machinery for implementing general order TDMVCC.

\abj{\section*{Supplementary material} \label{sec:supplementary_material}
The supplementary material contains information regarding the \ac{adga} computations performed for obtaining the benzoic acid \ac{pes} used in \revmads{Sec.~\ref{sec::num_results}}.
\revmads{It also contains an overview of all modes in the benzoic acid molecule, their harmonic frequencies, and whether they are treated at the TDMVCC or the TDH level in hybrid computations.}
}

\section*{Acknowledgements}
O.C. acknowledges support from the Independent Research Fund Denmark through grant number 1026-00122B.
All calculations were performed on the high performance computing cluster of the Center for Scientific Computing Aarhus. 
This work was supported by the Danish National Research Foundation through the Center of Excellence for Chemistry of Clouds (Grant Agreement No: DNRF172).

The authors thank Jonas Elm and the Independent Research Fund Denmark for financial support via grant number 9064-00001B.

\section*{Author Declarations}
\subsection*{Conflict of interest}
The authors have no conflicts to disclose.

\subsection*{Author contributions}
\textbf{Andreas Buchgraitz Jensen:} Conceptualization (equal); Data curation (lead); Formal analysis (equal); Investigation (lead); Software (lead); Visualization (lead); Writing/Original Draft Preparation (equal); Writing/Review \& Editing (equal).
\textbf{Mads Greisen H\o jlund:} Conceptualization (equal); Formal analysis (equal); Writing/Original Draft Preparation (equal); Writing/Review \& Editing (equal).
\textbf{Alberto Zoccante:} Conceptualization (equal); Formal analysis (equal); Writing/Original Draft Preparation (supporting); Writing/Review \& Editing (equal).
\textbf{Niels Kristian Madsen:} Conceptualization (equal); Formal analysis (equal); Software (supporting); Writing/Original Draft Preparation (supporting); Writing/Review \& Editing (equal).
\textbf{Ove Christiansen:} Conceptualization (equal); Formal analysis (equal); Project Administration (lead); Supervision (lead); Writing/Original Draft Preparation (equal); Writing/Review \& Editing (equal).

\section*{Data availability}
The data that support the findings of this study are available from the corresponding author upon reasonable request.

\bibliographystyle{jcp}

\begin{thebibliography}{10}

\bibitem{ple_routine_2023}
{\sc T.~Plé}, {\sc N.~Mauger}, {\sc O.~Adjoua}, {\sc T.~J. Inizan}, {\sc
  L.~Lagardère}, {\sc S.~Huppert}, and {\sc J.-P. Piquemal},
\newblock {\em J. Chem. Theory Comput.} {\bf 19}, 1432 (2023),
\newblock Publisher: American Chemical Society.

\bibitem{markland_nuclear_2018}
{\sc T.~E. Markland} and {\sc M.~Ceriotti},
\newblock {\em Nat. Rev. Chem.} {\bf 2}, 1 (2018),
\newblock Number: 3 Publisher: Nature Publishing Group.

\bibitem{christiansen_vcc_2004}
{\sc O.~Christiansen},
\newblock {\em J. Chem. Phys.} {\bf 120}, 2149 (2004).

\bibitem{hansen_timedependent_2019}
{\sc M.~B. Hansen}, {\sc N.~K. Madsen}, {\sc A.~Zoccante}, and {\sc
  O.~Christiansen},
\newblock {\em J. Chem. Phys.} {\bf 151}, 154116 (2019).

\bibitem{madsen_general_timedependent_2020}
{\sc N.~K. Madsen}, {\sc A.~B. Jensen}, {\sc M.~B. Hansen}, and {\sc
  O.~Christiansen},
\newblock {\em J. Chem. Phys.} {\bf 153}, 234109 (2020).

\bibitem{arponen1983vpl}
{\sc J.~Arponen},
\newblock {\em Ann Phys (N Y)} {\bf 151}, 311 (1983).

\bibitem{kvaal_ab_2012}
{\sc S.~Kvaal},
\newblock {\em J. Chem. Phys.} {\bf 136}, 194109 (2012).

\bibitem{pigg_time-dependent_2012}
{\sc D.~A. Pigg}, {\sc G.~Hagen}, {\sc H.~Nam}, and {\sc T.~Papenbrock},
\newblock {\em Phys. Rev. C} {\bf 86}, 014308 (2012).

\bibitem{hagen_coupled-cluster_2014}
{\sc G.~Hagen}, {\sc T.~Papenbrock}, {\sc M.~Hjorth-Jensen}, and {\sc D.~J.
  Dean},
\newblock {\em Rep. Prog. Phys.} {\bf 77}, 096302 (2014),
\newblock Publisher: IOP Publishing.

\bibitem{sverdrup_ofstad_time-dependent_nodate}
{\sc B.~S. Ofstad}, {\sc E.~Aurbakken}, {\sc {\O}.~S. Sch{\o}yen}, {\sc H.~E.
  Kristiansen}, {\sc S.~Kvaal}, and {\sc T.~B. Pedersen},
\newblock {\em Wiley Interdiscip. Rev. Comput. Mol. Sci.} , e1666 (2023),
\newblock eprint: https://onlinelibrary.wiley.com/doi/pdf/10.1002/wcms.1666.

\bibitem{pedersen_symplectic_2019}
{\sc T.~B. Pedersen} and {\sc S.~Kvaal},
\newblock {\em J. Chem. Phys.} {\bf 150}, 144106 (2019).

\bibitem{sato_communication:_2018}
{\sc T.~Sato}, {\sc H.~Pathak}, {\sc Y.~Orimo}, and {\sc K.~L. Ishikawa},
\newblock {\em J. Chem. Phys.} {\bf 148}, 051101 (2018).

\bibitem{nascimento_linear_2016}
{\sc D.~R. Nascimento} and {\sc A.~E. DePrince},
\newblock {\em J. Chem. Theory Comput.} {\bf 12}, 5834 (2016).

\bibitem{pedersen_interpretation_2021}
{\sc T.~B. Pedersen}, {\sc H.~E. Kristiansen}, {\sc T.~Bodenstein}, {\sc
  S.~Kvaal}, and {\sc {\O}.~S. Sch{\o}yen},
\newblock {\em J. Chem. Theory Comput.} {\bf 17}, 388 (2021).

\bibitem{wang_accelerating_2022}
{\sc Z.~Wang}, {\sc B.~G. Peyton}, and {\sc T.~D. Crawford},
\newblock {\em J. Chem. Theory Comput.} {\bf 18}, 5479 (2022),
\newblock Publisher: American Chemical Society.

\bibitem{madsen_time-dependent_2020}
{\sc N.~K. Madsen}, {\sc M.~B. Hansen}, {\sc O.~Christiansen}, and {\sc
  A.~Zoccante},
\newblock {\em J. Chem. Phys.} {\bf 153}, 174108 (2020).

\bibitem{hojlund_bivariational_2022}
{\sc M.~G. Højlund}, {\sc A.~B. Jensen}, {\sc A.~Zoccante}, and {\sc
  O.~Christiansen},
\newblock {\em J. Chem. Phys.} {\bf 157}, 234104 (2022).

\bibitem{hojlund_exponential_basis_2023}
{\sc M.~G. Højlund}, {\sc A.~Zoccante}, and {\sc O.~Christiansen},
\newblock {\em J. Chem. Phys.} {\bf 158}, 204104 (2023).

\bibitem{meyer_mctdh_1990}
{\sc H.-D. Meyer}, {\sc U.~Manthe}, and {\sc L.~Cederbaum},
\newblock {\em Chem. Phys. Lett.} {\bf 165}, 73 (1990).

\bibitem{beck_multiconfiguration_2000}
{\sc M.~Beck}, {\sc A.~Jäckle}, {\sc G.~Worth}, and {\sc H.-D. Meyer},
\newblock {\em Physics Reports} {\bf 324}, 1 (2000).

\bibitem{madsen_exponential_2018}
{\sc N.~K. Madsen}, {\sc M.~B. Hansen}, {\sc A.~Zoccante}, {\sc K.~Monrad},
  {\sc M.~B. Hansen}, and {\sc O.~Christiansen},
\newblock {\em J. Chem. Phys.} {\bf 149}, 134110 (2018).

\bibitem{ls_vscf09}
{\sc M.~B. Hansen}, {\sc M.~Sparta}, {\sc P.~Seidler}, {\sc O.~Christiansen},
  and {\sc D.~Toffoli},
\newblock {\em J. Chem. Theory Comput.} {\bf 6}, 235 (2010).

\bibitem{wan03:1289}
{\sc H.~Wang} and {\sc M.~Thoss},
\newblock {\em J.~Chem.\ Phys.} {\bf 119}, 1289 (2003).

\bibitem{wan09:024114}
{\sc H.~Wang} and {\sc M.~Thoss},
\newblock {\em J.~Chem.\ Phys.} {\bf 131}, 024114 (2009).

\bibitem{man08:164116}
{\sc U.~Manthe},
\newblock {\em J.~Chem.\ Phys.} {\bf 128}, 164116 (2008).

\bibitem{ven11:044135}
{\sc O.~Vendrell} and {\sc H.-D. Meyer},
\newblock {\em J.~Chem.\ Phys.} {\bf 134}, 044135 (2011).

\bibitem{wang_multilayer_2015}
{\sc H.~Wang},
\newblock {\em J. Phys. Chem. A} {\bf 119}, 7951 (2015).

\bibitem{selected_mctdh}
{\sc G.~A. Worth},
\newblock {\em J. Chem. Phys} {\bf 112}, 8322 (2000).

\bibitem{wodraszka_using_2016}
{\sc R.~Wodraszka} and {\sc T.~Carrington},
\newblock {\em J. Chem. Phys.} {\bf 145}, 044110 (2016).

\bibitem{wodraszka_systematically_2017}
{\sc R.~Wodraszka} and {\sc T.~Carrington},
\newblock {\em J. Chem. Phys.} {\bf 146}, 194105 (2017).

\bibitem{larsson_dynamical_2017}
{\sc H.~R. Larsson} and {\sc D.~J. Tannor},
\newblock {\em J. Chem. Phys.} {\bf 147}, 044103 (2017).

\bibitem{madsen_mrmctdh_2020}
{\sc N.~K. Madsen}, {\sc M.~B. Hansen}, {\sc G.~A. Worth}, and {\sc
  O.~Christiansen},
\newblock {\em J. Chem. Theory Comput.} {\bf 16}, 4087 (2020).

\bibitem{madsen_systematic_2020}
{\sc N.~K. Madsen}, {\sc M.~B. Hansen}, {\sc G.~A. Worth}, and {\sc
  O.~Christiansen},
\newblock {\em J. Chem. Phys.} {\bf 152}, 084101 (2020).

\bibitem{burghardt1999aat}
{\sc I.~Burghardt}, {\sc H.~Meyer}, and {\sc L.~Cederbaum},
\newblock {\em J. Chem. Phys.} {\bf 111}, 2927 (1999).

\bibitem{burghardt2008mqd}
{\sc I.~Burghardt}, {\sc K.~Giri}, and {\sc G.~A. Worth},
\newblock {\em J. Chem. Phys.} {\bf 129}, 174104 (2008).

\bibitem{worth_full_2003}
{\sc G.~A. Worth} and {\sc I.~Burghardt},
\newblock {\em Chem. Phys. Lett.} {\bf 368}, 502  (2003).

\bibitem{Richings15}
{\sc G.~Richings}, {\sc I.~Polyak}, {\sc K.~Spinlove}, {\sc G.~Worth}, {\sc
  I.~Burghardt}, and {\sc B.~Lasorne},
\newblock {\em Int. Rev. Phys. Chem.} {\bf 34}, 269 (2015).

\bibitem{shalashilin_multidimensional_2001}
{\sc D.~V. Shalashilin} and {\sc M.~S. Child},
\newblock {\em J. Chem. Phys.} {\bf 115}, 5367 (2001).

\bibitem{James16}
{\sc J.~A. Green}, {\sc A.~Grigolo}, {\sc M.~Ronto}, and {\sc D.~V.
  Shalashilin},
\newblock {\em J. Chem. Phys.} {\bf 144}, 024111 (2016).

\bibitem{curchod18}
{\sc B.~F.~E. Curchod} and {\sc T.~J. Mart{\'\i}nez},
\newblock {\em Chem. Rev.} {\bf 118}, 3305 (2018),
\newblock PMID: 29465231.

\bibitem{romer_gaussian-based_2013}
{\sc S.~Römer}, {\sc M.~Ruckenbauer}, and {\sc I.~Burghardt},
\newblock {\em J. Chem. Phys.} {\bf 138}, 064106 (2013),
\newblock Publisher: American Institute of Physics.

\bibitem{eisenbrandt_gaussian-based_2018}
{\sc P.~Eisenbrandt}, {\sc M.~Ruckenbauer}, and {\sc I.~Burghardt},
\newblock {\em J. Chem. Phys.} {\bf 149}, 174102 (2018),
\newblock Publisher: American Institute of Physics.

\bibitem{greene_tensor-train_2017}
{\sc S.~M. Greene} and {\sc V.~S. Batista},
\newblock {\em J. Chem. Theory Comput.} {\bf 13}, 4034 (2017),
\newblock Publisher: American Chemical Society.

\bibitem{dutra_quantum_2020}
{\sc M.~Dutra}, {\sc S.~Wickramasinghe}, and {\sc S.~Garashchuk},
\newblock {\em J. Chem. Theory Comput.} {\bf 16}, 18 (2020),
\newblock Publisher: American Chemical Society.

\bibitem{murakami_accurate_2018}
{\sc T.~Murakami} and {\sc T.~J. Frankcombe},
\newblock {\em J. Chem. Phys.} {\bf 149}, 134113 (2018),
\newblock Publisher: American Institute of Physics.

\bibitem{saller_quantum_2017}
{\sc M.~A.~C. Saller} and {\sc S.~Habershon},
\newblock {\em J. Chem. Theory Comput.} {\bf 13}, 3085 (2017),
\newblock Publisher: American Chemical Society.

\bibitem{baiardi_large-scale_2019}
{\sc A.~Baiardi} and {\sc M.~Reiher},
\newblock {\em J. Chem. Theory Comput.} {\bf 15}, 3481 (2019),
\newblock Publisher: American Chemical Society.

\bibitem{seidler2008tfc}
{\sc P.~Seidler}, {\sc M.~B. Hansen}, and {\sc O.~Christiansen},
\newblock {\em J. Chem. Phys.} {\bf 128}, 154113 (2008).

\bibitem{zoccante154101}
{\sc A.~Zoccante}, {\sc P.~Seidler}, and {\sc O.~Christiansen},
\newblock {\em J. Chem. Phys.} {\bf 134}, 154101 (2011).

\bibitem{christiansen2004sqf}
{\sc O.~Christiansen},
\newblock {\em J. Chem. Phys.} {\bf 120}, 2140 (2004).

\bibitem{MidasCpp}
{\sc O.~Christiansen}, {\sc D.~G. Artiukhin}, {\sc I.~H. Godtliebsen}, {\sc
  E.~M. Gras}, {\sc W.~Gy\H{o}rffy}, {\sc M.~B. Hansen}, {\sc M.~B. Hansen},
  {\sc M.~G. H\o{}jlund}, {\sc A.~B. Jensen}, {\sc N.~M. H\o{}yer}, {\sc E.~L.
  Klinting}, {\sc J.~Kongsted}, {\sc C.~K\"o{}nig}, {\sc D.~Madsen}, {\sc N.~K.
  Madsen}, {\sc K.~Monrad}, {\sc G.~Schmitz}, {\sc P.~Seidler}, {\sc
  K.~Sneskov}, {\sc M.~Sparta}, {\sc B.~Thomsen}, {\sc D.~Toffoli}, and {\sc
  A.~Zoccante},
\newblock MidasCpp.

\bibitem{sparta_adaptive_2009}
{\sc M.~Sparta}, {\sc D.~Toffoli}, and {\sc O.~Christiansen},
\newblock {\em Theor. Chem. Acc.} {\bf 123}, 413 (2009).

\bibitem{gfn2}
{\sc C.~Bannwarth}, {\sc S.~Ehlert}, and {\sc S.~Grimme},
\newblock {\em J. Chem. Theory Comput.} {\bf 15}, 1652 (2019),
\newblock PMID: 30741547.

\end{thebibliography}

\cleardoublepage
\listoffixmes
\end{document}


\title{Efficient time-dependent vibrational coupled cluster computations with time dependent basis sets
   at the two-mode coupling level: full and hybrid TDMVCC[2]. Supporting information}

\author{Andreas Buchgraitz Jensen}
\email{buchgraitz@chem.au.dk}
\affiliation{\au}

\author{Mads Greisen H\o jlund}
\affiliation{\au}

\author{Alberto Zoccante}
\affiliation{\upo}

\author{Niels Kristian Madsen}
\affiliation{\au}

\author{Ove Christiansen}
\email{ove@chem.au.dk}
\affiliation{\au}

\date{\today}

\maketitle

\section{ADGA settings for benzoic acid PES}
For generating the PES for benzoic acid through ADGA we used the following non-default settings:
\begin{itemize}
   \item \#1 Pes
   \begin{itemize}
      \item \#2 AnalyzeStates [39*3]
      \item \#2 SymThr: 5.0e-4
   \end{itemize}
   \item \#1 Vib
   \begin{itemize}
      \item \#2 Basis
      \begin{itemize}
         \item \#3 ScalBounds: 1.5 30.0
         \item \#3 UseScalingFreqs
         \item \#3 MaxNoPrimBasisFunctions: 750
      \end{itemize}
   \end{itemize}
\end{itemize}

\section{Table of modes included in TDMVCC[2|X]}

\begin{longtblr}[
   caption = {Harmonic frequencies for all modes and our inclusion of modes in TDMVCC[2|X]},
   label = {tab::included_modes},
]{
   width = 0.95\linewidth, colspec = {|X[1]|X[3]|X[1,c]|X[1,c]|X[1,c]|},
   rowhead = 1,
   hlines, vlines,
}
Mode & Harmonic frequencies [cm$^{-1}$] & X = 8 & X = 13 & X = 18 \\
0 & 49.35192887375352 &  &  & $\checkmark$ \\
1 & 144.21336276881505 &  &  & $\checkmark$ \\
2 & 191.8173174260282 &  &  & $\checkmark$ \\
3 & 349.38132696407865 &  &  &  \\
4 & 368.4487303565827 &  &  & $\checkmark$ \\
5 & 396.13741103420887 & $\checkmark$ & $\checkmark$ & $\checkmark$ \\
6 & 477.43416741763156 &  &  & $\checkmark$ \\
7 & 587.4260556695822 &  &  &  \\
8 & 596.205522248642 & $\checkmark$ & $\checkmark$ & $\checkmark$ \\
9 & 604.694981037908 & $\checkmark$ & $\checkmark$ & $\checkmark$ \\
10 & 652.8376118806012 &  & $\checkmark$ & $\checkmark$ \\
11 & 701.8277763585781 &  & $\checkmark$ & $\checkmark$ \\
12 & 768.8009650135224 & $\checkmark$ & $\checkmark$ & $\checkmark$ \\
13 & 785.5962736213646 &  &  &  \\
14 & 879.5756943503557 &  &  &  \\
15 & 920.0659395599141 &  &  &  \\
16 & 931.1616419951641 &  &  &  \\
17 & 939.4276055260843 &  &  &  \\
18 & 981.3648916614968 &  &  &  \\
19 & 1077.850754261199 &  &  &  \\
20 & 1102.5320754289394 & $\checkmark$ & $\checkmark$ & $\checkmark$ \\
21 & 1124.2580258203127 &  &  &  \\
22 & 1146.8939538220052 & $\checkmark$ & $\checkmark$ & $\checkmark$ \\
23 & 1181.0843894622378 &  &  &  \\
24 & 1191.464517627389 &  & $\checkmark$ & $\checkmark$ \\
25 & 1271.119631020361 &  & $\checkmark$ & $\checkmark$ \\
26 & 1277.5958533834605 &  & $\checkmark$ & $\checkmark$ \\
27 & 1329.2755215725035 &  &  &  \\
28 & 1435.0540982412303 &  &  &  \\
29 & 1462.0683302395003 &  &  &  \\
30 & 1585.9986505130387 &  &  &  \\
31 & 1593.7717203967077 &  &  &  \\
32 & 1756.0482572469946 & $\checkmark$ & $\checkmark$ & $\checkmark$ \\
33 & 3078.6432404182538 &  &  &  \\
34 & 3085.864890062676 &  &  &  \\
35 & 3093.293452297898 &  &  &  \\
36 & 3101.600220035311 &  &  &  \\
37 & 3108.3122532680463 &  &  &  \\
38 & 3458.4957352378897 & $\checkmark$ & $\checkmark$ & $\checkmark$ 
\end{longtblr}